\documentclass[12pt]{article}

\usepackage[utf8]{inputenc}
\usepackage{bm}
\usepackage{graphicx}
\usepackage{pdfpages}
\usepackage{bbold}
\usepackage{enumitem}
\usepackage{tikz}
\usepackage[margin=1in]{geometry} 
\usepackage{setspace} 
\usepackage[english]{babel}
\RequirePackage[colorlinks,citecolor=blue,urlcolor=blue]{hyperref}
\usepackage{hyperref} 

\usepackage{csquotes}

\usepackage{setspace}
\usepackage[citestyle=authoryear, uniquename=false, uniquelist=false, bibstyle=authoryear, maxcitenames = 2, maxbibnames = 99, url=false, natbib]{biblatex}

\AtEveryBibitem{\clearfield{month}}
\AtEveryBibitem{\clearfield{day}}

\renewbibmacro{in:}{}
\AtEveryBibitem{%
  \clearlist{language}%
}

\DeclareNameFormat{family}{%
  \usebibmacro{name:family}
    {\namepartfamily}
    {\namepartgiven}
    {\namepartprefix}
    {\namepartsuffix}%
  \usebibmacro{name:andothers}}

\DeclareNameAlias{default}{family}

\DeclareNameAlias{sortname}{default}
\DeclareNameAlias{labelname}{default}

\usepackage{amsmath} 
\usepackage{bbm} 
\usepackage{bm}
\usepackage{amssymb}
\usepackage{multirow} 
\usepackage{units} 
\usepackage{xfrac} 

\DeclareMathOperator{\E}{\mathbb{E}} 
\DeclareMathOperator*{\argmin}{argmin}
\newcommand{\indep}{\rotatebox[origin=c]{90}{$\models$}}
\newcommand{\BEA}{\begin{eqnarray*}}

\newcommand{\EEA}{\end{eqnarray*}}
\newcommand{\BE}{\begin{enumerate}}
\newcommand{\EE}{\end{enumerate}}
\newcommand{\eps}{\varepsilon}
\newcommand{\bmat}{\begin{pmatrix}}
\newcommand{\emat}{\end{pmatrix}}
\newcommand{\bbmt}{\begin{bmatrix}}
\newcommand{\ebmt}{\end{bmatrix}}
\newcommand{\bpmt}{\begin{pmatrix}}
\newcommand{\epmt}{\end{pmatrix}}

\newcommand{\bp}{\begin{proof}}
\newcommand{\ep}{\end{proof}}

\newcommand{\BEABC}{\begin{enumerate}[label=(\alph*)]}

\usepackage{fancyhdr}
\fancypagestyle{firstpage}{
  \fancyhf{}
  \fancyhead[R]{}
  \fancyhead[L]{}
  \fancyhead[C]{This paper is published with open access in \textit{The American Statistician} \url{https://doi.org/10.1080/00031305.2023.2257237}}
  \fancyfoot[C]{\thepage}
}

\usepackage{cleveref}[2012/02/15]
\crefformat{footnote}{#2\footnotemark[#1]#3}

\title{Statistical Challenges in Online Controlled Experiments: A Review of A/B Testing Methodology}
\author{Nicholas Larsen\footnote{North Carolina State University, Department of Statistics}, Jonathan Stallrich\footnotemark[\value{footnote}], Srijan Sengupta\footnotemark[\value{footnote}], Alex Deng\footnote{Airbnb}, \\Ron Kohavi, Nathaniel T. Stevens\footnote{University of Waterloo, Department of Statistics \& Actuarial Science, \url{nstevens@uwaterloo.ca}}}

\date{}

\begin{document}

\maketitle
\thispagestyle{firstpage}
\begin{abstract}

The rise of internet-based services and products in the late 1990's brought about an unprecedented opportunity for online businesses to engage in large scale data-driven decision making. Over the past two decades, organizations such as Airbnb, Alibaba, Amazon, Baidu, Booking.com, Alphabet's Google, LinkedIn, Lyft, Meta's Facebook, Microsoft, Netflix, Twitter, Uber, and Yandex have invested tremendous resources in \textit{online controlled experiments} (OCEs) to assess the impact of innovation on their customers and businesses. Running OCEs at scale has presented a host of challenges requiring solutions from many domains. In this paper we review challenges that require new statistical methodologies to address them. In particular, we discuss the practice and culture of online experimentation, as well as its statistics literature, placing the current methodologies within their relevant statistical lineages and providing illustrative examples of OCE applications. Our goal is to raise academic statisticians' awareness of these new research opportunities to increase collaboration between academia and the online industry.


\end{abstract}

\newpage


\section{Introduction}

\subsection{Background}\label{sec:background}

It is estimated that in 2022 globally, 5.16 billion people (64.4\% of the world's population) used the internet, each engaging with it on average 6.5 hours per day, and in aggregate spending over \$5 trillion (USD) on consumer goods, travel and tourism, digital media, and health-related products and services \citep{digital2023}. In 2023, e-commerce is predicted to account for 21\% of all commerce, and by 2025 that number is expected to grow to nearly 25\% \citep{ecommerce2022}. Given this scale of internet use, it is unsurprising that the optimization of online products and services is of great interest to online businesses and online components of traditional brick-and-mortar businesses.

\textit{Online controlled experiments} (OCEs), digital versions of randomized controlled trials (RCTs)\citep{box_hunter_hunter_2005}, provide a means to do this; OCEs seek to use user-generated data to test and improve internet-based products and services \citep{Kohavi2023}. Informally referred to as \textit{A/B tests}, OCEs are an indispensable tool for major technology companies when it comes to maximizing revenue and optimizing the user experience \citep{luca2021power}. Industry giants run hundreds of experiments on millions of users every day \citep{gupta_summit_2019}, testing changes along multiple axes including: websites, services, and installed software; desktop and mobile devices; front- and back-end product features; personalization and recommendations; and monetization strategies. With OCEs, the causal impact of such changes--whether it be positive, negative, or zero--can be estimated. We acknowledge that observational methods for causal inference are also relevant here, though their use is much less prominent. We discuss them briefly in Section SM4 of the Supplementary Material.

While most positive changes are small, and improvement is incremental \citep{bojinov2022online}, results from OCEs have the potential to be incredibly lucrative. Google's famous ``41 shades of blue" experiment is a classic example of an OCE that translated into a \$200 million (USD) increase in annual revenue \citep{Hern_why_google}; Amazon used insights from an OCE to move credit card offers from the homepage to the checkout page, resulting in tens of millions (USD) in profit annually \citep{kohavi_surprising_2017}; Bing deployed an A/B test for ad displays that resulted in \$100 million (USD) of additional revenue in the U.S. alone \citep{kohavi_trustworthy_2020}.  Even though such million-dollar ideas are relatively rare, the net gains from OCEs have been so profound that many organizations have completely overhauled their business models, with experimentation at the epicenter \citep{thomke2020experimentation}. For instance, Netflix attributes its membership growth from 2 countries to over 190 in the span of just 6 years to its adoption of online controlled experimentation \citep{all_about_ab_testing}, and Duolingo's 2022 Q2 shareholder letter attributes their growth to an ``A/B test everything'' mentality \citep{duolingo2022testeverything}. The document even includes a description of their A/B testing process and several examples of how the product as evolved through experimentation.

Organizations that have accepted OCEs as standard practice generally adopt a so-called ``culture of experimentation," which is rooted in three tenets \citep{kohavi_online_2013}: (1) the organization wants to make data-driven decisions, (2) the organization is willing to invest in the people and infrastructure needed to run trustworthy experiments, and (3) the organization recognizes that it is poor at assessing the value of ideas. Generally, more than 50\% of ideas fail to generate meaningful improvements \citep{kohavi_trustworthy_2020}. And in some domains, the failure rate of experiments (due to a combination of bad ideas or buggy implementation) is 90\% or higher \citep{kohaviIntuitionBusters}. Thus, \textit{carefully} executed experiments provide a trustworthy, data-driven means to determine which ideas improve key metrics, which hurt, and which have no detectable impact, allowing the organization to invest in those that work, while pivoting to avoid the others. Within this culture, the attitude of ``\textit{more, better, faster}" is prevalent \citep{tang_overlapping_2010}; organizations strive to increase the number of experiments so that all changes are properly evaluated; invalid experiments and harmful combinations of variants are straightforward to identify; and deployment, run time, and analysis occur within a relatively short period of time. 

Compared to physical controlled experiments (in e.g., agriculture, manufacturing, pharmaceutical development), the cost incurred to design and run an OCE is low, even negligible for organizations with expertise in software development and statistics. Consequently, practitioners are able to run large numbers of experiments with potentially enormous sample sizes. In the case of large tech organizations, the combination of new features and modifications can result in billions of different versions of a given product \citep{kohavi_online_2013}. In these cases, hundreds of thousands of users are randomized concurrently to hundreds of experiments running simultaneously, so users may be in hundreds of experiments at a time \citep{gupta_summit_2019}. Companies performing OCEs at this scale typically use \textit{experimentation platforms} (software that is licensed or developed in-house) to automate the experimentation process, such as randomizing users, collecting data, managing concurrent experiments, and generating analysis reports \citep{microsof_exp_platform, tang_overlapping_2010, linkedin_exp_platform, fabijan_exp_platform_survey}. See \citet{inhouse_exp_platform} for a catalogue of in-house experimentation platforms developed by several prominent tech companies. Smaller companies tend to opt for third-party vendors that specialize in setting up, deploying, and analyzing OCEs. Several vendors were compared in \citet{kohavi2023slides} including (in alphabetical order): A/B Smartly, AB Tasty, EPPO, GrowthBook, Kameleoon, Optimizely, Split, Statsig, VWO, and Webtrends-Optimize. In all cases, this level of automation necessitates data quality checks like A/A tests and sample ratio mismatch (SRM) tests to establish trust in the experimentation platform. (For further discussion of these practices and challenges, see Chapters 19 and 21 of \cite{kohavi_trustworthy_2020}, and the introduction in \cite{lindon_anytime_2020}.) 

In this online setting, with the culture of testing as many ideas as possible, as quickly as possible, novel practical issues and modern challenges abound (see, e.g., \citet{gupta_summit_2019,bojinov2022online,quin2023b} for nontechnical discussions, and \citet{georgiev2019statmethods} for a technical primer). The context in which OCEs operate departs markedly from the original applications for which traditional experiments were developed nearly a century ago; understanding this context is vital for developing relevant methodology for OCEs. For statisticians, online controlled experimentation provides a host of new opportunities for methodological and theoretical development. New approaches that fit the nuances of OCE applications are in high demand, with the majority of cutting-edge research spearheaded by those in industry. The purpose of this paper, therefore, is to review the statistical methodology associated with OCEs, summarize its accompanying literature, and provide an overview of open statistical problems. We hope that increasing academic statisticians' awareness of these research opportunities will help to bridge the gap between academia and the online industry. 

\subsection{The General Framework} \label{sec:notation}

Here we introduce the notation and key terms that will be used throughout this review and we describe the basic statistical framework for OCEs. It is useful to note that as a field, online experimentation has developed disparately across industries and domains, thus there are no unifying standards with respect to methodological approach and notation; even the term ``controlled experiment" goes by different names depending on the organization: ``flights" at Microsoft, ``bucket tests" at Yahoo, ``field experiments" at Facebook, and ``1\% experiments/click evaluations" at Google. Standard conventions would bring useful unification to this field. The following notation largely draws from traditional RCT and causal inference literature, and is intended to help unify much of the OCE literature.

Let $K$ be the number of variants (also known as buckets, arms, splits, and treatments) that compose the experiment. Ordinarily one of these variants is a control against which all other variants are compared. Unless explicitly stated, we shall assume for the rest of this review a standard treatment-versus-control setup, in which case $K = 2$. While multi-variant ($K>2$) experiments exist in this space (colloquially referred to as ``A/B/n tests''), we focus on the $K=2$ ``A/B test" for pedagogical reasons; even with $K>2$ variants, determining which is optimal typically reduces to a pairwise comparison between each treatment and the control.

In such experiments, $n$ experimental units (e.g., users, cookies, sessions, etc.) are typically randomized in real time to one of the variants, and a response observation $Y_i$ is collected for each $i=1,\ldots,n$. It is important to note that these response observations are typically themselves aggregates of more granular raw event data \citep{metric_comp2020}. For instance, consider the response variable \textit{number of clicks per user} which may be a count per user aggregated across sessions and/or pages. Interest then lies in optimizing some \textit{metric}, which is a numerical summary of the response. Extending the previous example, interest may lie in maximizing the \textit{average} number of clicks per user. Such metrics are often, but not always, averages. In some contexts, quantile or double-average metrics may be more suitable. We discuss such applications in more detail in Section SM4 of the Supplementary Material. 

For simplicity of exposition, we have described a situation with one metric and hence one response variable. However, in practice there may be hundreds (even thousands) of metrics computed, many of which are used for debugging, some of which may be organizational \textit{guardrail} metrics that the experimenters wish to avoid negatively impacting, and a small number of which compose the \textit{overall evaluation criterion (OEC)} which is to be optimized. In general, defining and selecting metrics (as well as their corresponding randomization and analysis units) are key components of OCEs and we direct the reader to \citet{crook_seven_2009, deng_data-driven_2016, dmitriev_dirty_2017, kohavi_trustworthy_2020} for further discussion. 

When the metric is an average, the primary goal of the experiment is to estimate the \textit{average treatment effect (ATE)}; the difference between the average outcome when the treatment is applied globally and when the control is applied globally. Within the potential outcomes framework \citep{neyman1923application,rubin1974estimating}, $Y_i(0)$ represents unit $i$'s response in the hypothetical scenario where $i$ receives the control, and $Y_i(1)$ is the potential response when unit $i$ receives the treatment. Letting $W_i$ denote the binary treatment indicator for unit $i$, and given a particular treatment assignment to all experimental units $\mathbf{W} = \big(W_1, \dots , W_n \big)'$, the expected outcome is
$\E\big[\frac{1}{n}\sum_{i=1}^nY_i(W_i)\big] = \mu(\mathbf{W})$, and the ATE is therefore given by 
\begin{equation}
\label{eqn:ATE} 
\begin{aligned}
\tau &= \mu(\mathbf{1}) - \mu(\mathbf{0}) \\
&= \frac{1}{n}\sum_{i=1}^n\E[Y_i(1) - Y_i(0)],
\end{aligned}
\end{equation}
where these expectations may be taken with respect to the random sampling or the random assignment, depending on whether inference is sample-based or design-based \citep{abadie2020sampling}. In reality, $i$ can only be assigned to a single variant at a time, thus one cannot directly observe both $(Y_i(0), Y_i(1))$ and so the ATE is typically estimated with the difference-of-group-means estimator,
\begin{equation}
    \hat{\tau} = \frac{1}{n_1}\sum_{\{i:W_i=1\}}Y_i - \frac{1}{n_0}\sum_{\{i:W_i=0\}}Y_i,
\end{equation}
where $n_0$ and $n_1$ are respectively the sizes of the control and treatment groups such that $n_0 + n_1 = n$. In practice it is also common to define the treatment effect as a relative percent, often referred to as \textit{lift}, since it is easier to interpret and it is more stable (over experiment duration, for example). 

Statistical significance is the most common mechanism by which a given treatment's effectiveness is affirmed in an A/B test. Analyses of A/B tests are therefore most often carried out via two-sample hypothesis tests for $\tau$ with standard test statistics of the form $\hat{\tau}/\hat{\sigma}_{\hat{\tau}}$. Such analyses, and the designs that generate data for them, commonly assume that the response of each user does not depend on other users' treatment assignments (the \textit{Stable Unit Treatment Value Assumption}, or SUTVA). SUTVA is a reasonable assumption for many scenarios; however in Section \ref{interference} we discuss OCE settings where the assumption is violated and alternative methodologies are necessary. In many scenarios, sample sizes are large enough to confidently exploit the central limit theorem, permitting the use of the standard normal null distribution. There are, however, scenarios in which only a fraction of the user base is experimented on and asymptotic normality cannot be assumed. Such scenarios are discussed in Section \ref{sec:triggering}. Given the heavy reliance on p-values it is important to acknowledge that the reproducibility crisis stemming from the misuse of hypothesis tests also plagues OCEs; p-value misinterpretation and problematic practices regularly lead to increased false-positive rates \citep{berman2021false, kohaviIntuitionBusters}. This is an area of ongoing practical and methodological concern in many fields, including online experiments.

\subsection{Roadmap} \label{sec:layout}

With this context and foundation laid, we now review the statistical research in this area and discuss the many open problems. The article proceeds as follows. Section \ref{sec:sensitivity_and_small_effects} discusses techniques for improving experimental power -- a critical issue despite the relatively large sample sizes in OCEs. Sections \ref{sec:hetero_effects} and \ref{sec:long_term_effects} respectively present literature regarding the challenges of estimating heterogeneous and long-term treatment effects. Section \ref{sec:optstop} discusses the problem of optional stopping and how sequential testing methods have been adapted to run online experiments. All of these sections presume SUTVA holds; we summarize the literature that explores violations of this assumption in Section \ref{interference}. We conclude the review with a call to action for further collaboration between academia and OCE practitioners in Section \ref{sec:the_end}. Note that a Supplementary Material file accompanies the paper in which we provide expanded coverage of certain topics from the main text, as well as a brief discussion of additional topics outside the scope of this review.


\section{Sensitivity and Small Treatment Effects}
\label{sec:sensitivity_and_small_effects}

\noindent \textbf{Motivating Example:} \textit{Suppose an e-commerce website observes that 5\% of their visitors make a purchase and the average purchase is \$25 with a standard deviation of $\sigma=\$6$ during a one-week experiment period. Therefore, on average, visitors spend \$1.25. Suppose also the company's annual revenue is \$20 million, and gains or losses of \$1 million (5\%) are material. If the company wishes to run an experiment and detect a 5\% change in revenue (i.e., $\delta = 1.25\times0.05$) with 80\% power at a 5\% significance level, a rough sample size calculation indicates they need $n_0=n_1=16\sigma^2/\delta^2 = (16\times6^2)/(1.25\times0.05)^2=147,456$ users per variant. This is reasonable for a small startup. However, suppose now that the company's annual revenue is \$50 billion, with gains or losses of \$10 million (0.02\%) of interest. An experiment designed to detect a 0.02\% change in revenue (i.e., $\delta = 1.25\times0.0002$) requires $n_0=n_1=16\sigma^2/\delta^2 = (16\times6^2)/(1.25\times0.0002)^2=9.2$ billion users per variant, i.e., 18.4 billion users in a single week. The human population on Earth is about 8 billion at the time of writing, so it is impossible for this company to detect changes that would gain or lose them \$10 million per year.}

Many leading organizations at the forefront of online controlled experimentation have user populations numbering in the hundreds of millions, if not billions. However, the sentiment that OCEs do not suffer from inadequate sample sizes is misconceived \citep{tang_overlapping_2010}. Given the fundamental relationship between sample size and an experiment's ability to detect true, nonzero treatment effects (namely its power), a key challenge facing even the largest of organizations is designing adequately powered experiments. A naive solution would be to simply extend the experiment's duration, thereby increasing the number of users. However, as we elaborate upon in Section \ref{sec:long_term_effects}, this practice is ill-advised. Instead, it is better practice to employ a tactic that is tailored to the reason for insufficient power, which is generally one of three causes. 

First, the treatment impacts the entire user population and the effect is roughly homogeneous, but very small in magnitude. As illustrated in the opening examples, even a fraction of a percent-change can translate to millions of dollars in revenue. We discuss the literature around this issue in Section \ref{sec:transforming_Y}. Second, many experiments test features that do not affect all users, making the treatment effect highly attenuated (Section \ref{sec:triggering}). Third, the treatment effects on known subpopulations are of interest, where sample sizes are smaller by definition (we defer this discussion to Section \ref{sec:hetero_effects}). In general, research regarding improving experimental power for OCEs tends to focus on boosting \textit{sensitivity}, either by directly reducing the variance of $Y_i$ or by reducing the variance of estimators for $\tau$. The aforementioned subsections provide an overview of common methodology in this area. While specific methods of combating inadequate power are reviewed in this section, we encourage the reader to keep in mind that the issue of adequate power applies to all the challenges subsequently discussed in this review.

\subsection{Transforming $Y$, Method of Control Variates, and Stratified Sampling}
\label{sec:transforming_Y}

In order to improve sensitivity, a common approach is to transform $Y$ into $Y^*$ of lower variance which, all else being equal, translates to a lower variance estimator of $\tau$. In online experiments there can be dozens, even hundreds of metrics of potential interest, many with different properties that make it all but impossible to identify a ``one size fits all" transformation. Much work has been devoted to documenting metric behavior and discussing techniques for metric definition. \citet{kohavi_seven_2014} describe several examples of non-intuitive metric behavior and other peculiarities, illustrating the benefits of identifying skewed metrics and capping (truncating) them to improve sensitivity. Other transformations for improving the sensitivity of $Y$ include 
binarizing count metrics and revenue. \citet{deng_data-driven_2016} define \textit{directionality} (consistent behavior in one direction for positive treatment effects and in the opposite direction for negative effects) as an important feature when choosing metrics, suggesting that one should leverage prior experiments to compile a corpus of good metrics and to evaluate sensitivity and directionality with Bayesian priors. \citet{deng_data-driven_2016} also propose aggregating metrics in the form of a weighted linear combination, which is adopted and expanded upon in \citet{kharitonov2017_learning}. They frame finding sensitive combinations of metrics as a machine learning problem, incorporating both labeled and unlabeled data from past experiments. In \citet{drutsa2015future}, features are extracted from data while the experiment is running and used to forecast metrics over a hypothetical post-experiment period. The authors also note their methodology may be applied to long-term effect estimation using statistical surrogacy, which we further discuss in Section \ref{sec:long_term_effects}.

In addition to transformations of $Y$, a popular approach is to define an efficient, mean-zero augmented estimator of $\tau$ using the method of control variates \citep{cuped_towardsdatascience, cuped_statsig, Bending_time}. Briefly, this method assumes, in addition to i.i.d. $\{Y_i\}_{i=1}^n$, the availability of independent observations of a covariate, $\{X_i\}_{i=1}^n$, such that $\E[X_i] = \mu_x$. 
Often, these covariate measurements are collected from prior logs or experiments. Let $Y_i^* = Y_i - \theta (X_i - \mu_x)$, then $$Var(Y_i^*) = Var(Y_i) + \theta^2Var(X_i) - 2\theta Cov(Y_i,  X_i)$$ is minimized with respect to $\theta$ using the OLS solution $Cov(Y_i, X_i)/Var(X_i)$. Putting this together in the context of sample means gives $$Var(\overline{Y}^*) =  (1 - \rho^2)Var(\overline{Y}) \leq Var(\overline{Y}),$$ where $\rho = Corr(Y_i,X_i)$. Thus, an ATE estimator that uses the difference of treatment and control means of $Y^*_i$ tends to have lower variance than the traditional $\hat{\tau}$, particularly when $X_i$ is strongly correlated with $Y_i$. For OCEs, this technique is referred to as CUPED (Controlled experiments Utilizing Pre-Experiment Data) and was first proposed by \citet{deng_improving_2013}. The authors empirically demonstrate that an effective covariate choice is the same variable $Y_i$ but collected during a pre-experiment period ($X_i \equiv Y_i^{\text{pre}}$). Such a choice can drastically increase sensitivity and thereby reduce time to statistical significance in determining $H_1: \tau \neq 0$. The authors also demonstrate that $\mu_x$ need not be known when $X_i$ is uncorrelated with $W_i$ and they also emphasize that despite resembling ANCOVA, CUPED does not require any linear model assumptions and can be treated as efficiency augmentation as in semi-parametric estimation \citep{tsiatis2006semiparametric}. Consequently, CUPED has become a standard tool for many practitioners, although it is important to note that it can only be applied to users for which prior information exists \citep{gupta_summit_2019, CUPED_at_BBC, drusta2015practical, HowBook_online, evidence_of_CUPED_at_airbnb}. 

A key open question with respect to CUPED applications concerns the situation when the covariate alone is not sufficiently correlated with the response. An approach that shows promise employs synthetic controls, where one identifies a similar population without treatment exposure to use as covariates for modeling $Y$ \citep{zhang2021regression}. Another technique is to take advantage of a phenomenon that occurs in online experiments known as ``triggering" \citep{deng2023zero}, which we further discuss in Section \ref{sec:triggering}. Further research with respect to the interplay between CUPED and other standard variance reduction techniques is also of interest. \citet{xie_improving_2016} apply CUPED to large-scale A/B tests for a subscription streaming service, and \citet{liou2020variance} compare CUPED against variance-weighted estimators on a social media platform, finding that an aggregation of the two methods outperformed either individually. \citet{deng_improving_2013} note that CUPED also permits nonlinear adjustments to the response variable. Following this, \citet{poyarkov2016boosted} develop an approach that assumes each user has a response $Y$ and a set of features $\mathbf{F} \in \mathbb{R}^p$ independent of treatment assignment. Let $Y = f(\mathbf{F})$, where $f$ is an unknown, non-parametric function that is estimated with machine learning. Following the general idea of control variates, the covariate is chosen to be the predicted outcomes of $\hat{f}$. \citet{poyarkov2016boosted} then use $Y^* = Y - \hat{f}(\mathbf{F})$ as the primary metric for estimating the ATE, noting an increase in sensitivity compared to traditional A/B tests.

Closely related to the method of control variates/CUPED is stratified sampling. We discuss these connections as well as the use and drawbacks of stratified sampling in more detail in Section SM1 of the Supplementary Material.


\subsection{Triggered Analysis}
\label{sec:triggering}

\noindent \textbf{Motivating Example:} \textit{Suppose engineers are testing a change made on an e-commerce website's checkout page. Users in the experiment who never interact with this checkout page are not impacted by the experiment and so their treatment effect is zero. Many such users will increase noise and dilute the treatment effect. Sensitivity may be increased by analyzing only the users who could have been impacted by the experiment; those that were triggered into the analysis. Although this reduces sample size, the treatment effect among the triggered users is undiluted and therefore higher and easier to detect.}

\textit{Triggered analysis} broadly refers to an OCE analysis where one only considers users who have the potential of being impacted by an experiment, excluding those who would not be affected by the proposed variant \citep{deng2023zero, kohavi_controlled_2009, kohavi_online_2012, xu_sqr_2018}. Users are said to have \textit{triggered} the experiment when they exhibit behavior that results in direct exposure to their assigned variant.  Such users could be assigned to a variant either early (e.g., when first visiting the web site) or when they exhibit some type of behavior that triggers the experiment (e.g., when reaching the checkout page, which impacts the variant they receive). Key analysis challenges include: (1) generalizing the results from the triggered users to a broader population, and (2) reducing the variance of $\tau$ estimators to offset the smaller sample sizes that result from triggering. For an in-depth discussion of triggering case-studies, including the example above, see Chapter 20 of \citet{kohavi_trustworthy_2020}. In what follows, we provide a brief description and an overview of research in this area.

Let $\Omega$ be the overall user population and $\Theta \subset \Omega$ the population of users who could be affected by the treatment. A given user is determined to belong to $\Theta$ via techniques such as conditional checks or counterfactual logging \citep{kohavi_trustworthy_2020, deng2023zero}. If $\Theta$ comprises only a modest fraction of $\Omega$, (i.e., $\frac{|\Theta|}{|\Omega|} \leq 0.2$, for instance), an experiment that samples data from the entire population could be severely under-powered, particularly when effect sizes are small \citep{kohavi_controlled_2009}. To mitigate this issue, practitioners focus analysis only on triggered users. The difference-of-means estimator $\hat{\tau}_\Theta$ is an unbiased estimator for the ATE of the triggered population, $\tau_\Theta$, under standard assumptions. However, $\tau_\Theta$ is typically larger than the population-level $\tau_\Omega$ and the corresponding estimator generally has greater variance. The process of estimating $\tau_\Omega$ with $\hat{\tau}_\Theta$ is referred to as estimating the \textit{diluted treatment effect}. This is of interest because it allows the business to estimate the global impact of launching a new feature to all users (some of whom will be directly impacted, and some of whom will not). Below we describe methodological development in this area.

Most triggered analyses fall under the following framework. Assume a (not necessarily random) sample of $N$ units, $n$ of which are triggered. Each user $i$ interacts with the website on multiple separate events. During each event, $i$ may or may not trigger (e.g., $i$ may interact with the checkout page during one event, but not the other). The most common analysis technique is the \textit{user-trigger analysis}, which incorporates all events beginning with the first event where $i$ triggered. Such analyses are quite popular as they do not require any assumptions regarding the treatment effect, and are amenable to common user-level metrics. \citet{chen2018automatic} utilize the user-trigger framework to illustrate the benefits of triggered analyses in terms of power gains and variance reduction, as well as to highlight the types of biases that may occur under such approaches. The \textit{session-trigger analysis} is another approach that groups events into ``sessions" and only keeps sessions that contain trigger events for analysis. While \citet{deng2015diluted} note that estimates from session-triggered analyses do tend to have lower variance than user-trigger analyses, the treatment effect must be zero in the nontriggered sessions in order for this approach to be valid. While perhaps true in some cases, generally this assumption is difficult to verify for most applications. 

One approach for estimating the diluted treatment effect is to derive $\tau_\Omega$ in terms of $\tau_\Theta$, producing so-called ``diluted formulas". For additive metrics, $Y_i = Y_{i,t} + Y_{i,u}$, where $Y_{i,t}$ is an outcome when $i$ is triggered and $Y_{i,u}$ for when $i$ is untriggered, it can be shown that the diluted treatment effect is the average treatment effect on triggered users weighted by the proportion of triggered users, i.e., $\tau_\Omega = \tau_\Theta \times \frac{n}{N}$. Note that this expression only applies when, for $i \in \Theta$, there is no treatment effect on the sessions where $i$ is untriggered, i.e., $Y_{i,u}(1) - Y_{i,u}(0) = 0$. In other words, this expression is only for valid session-trigger analyses. With ratio metrics, $Y_i = \frac{a_i}{b_i}$, if there is no treatment effect for the denominator term, $b_i = b_i(1) = b_i(0)$, and the rate at which users are triggered into the experiment is independent of $\tau_\Theta$, then the diluted formula is $\tau_\Omega = \tau_\Theta \times \frac{n}{N} \times \Bar{r}$, where $\Bar{r}$ is the average trigger rate as a function of $b_i$. Further details for these derivations may be found in \citet{deng2015diluted}, but it is important to note that these formulas only apply when the treatment effect defined in \eqref{eqn:ATE} is of interest. When a relative effect (i.e., lift) is of interest, one must perform relative dilution where weighting is not by the proportion of triggered users, but by their contribution to the metric. While these formulas are certainly helpful in illustrating the connection between $\tau_\Theta$ and $\tau_\Omega$, they are restrictive because their underlying assumptions are not necessarily realistic and closed-form expressions only exist for special cases. As noted by \citet{deng2015diluted}, the trigger rates are rarely independent of the triggered treatment effect; users who visit a website frequently will have higher trigger rates and tend to have a larger treatment effect than less-frequent users (see \citet{wang2019heavy}). 

The above formulas are provide a solution for estimating $\tau_\Omega$ but they do not address the issue of low power that typically afflicts triggered analyses. \citet{deng2015diluted, deng2023zero} simultaneously address both issues by formalizing the connection between all diluted formulas and variance reduction. Under the assumption that there is no treatment effect when users are untriggered, \citet{deng2015diluted} apply the CUPED framework (Section \ref{sec:transforming_Y}) by augmenting $\hat{\tau}_\Theta$ with mean-zero data from the trigger complement group. The authors show that the resulting augmented estimator is unbiased for $\tau_\Omega$, can achieve appreciable variance reduction, and applies to metrics of any form. \citet{deng2023zero} extend this application of CUPED to one-sided triggering, a type of one-sided noncompliance where only the triggering status of the treatment group is observed.

Compared to other challenges in OCEs, the literature for triggering is, at present, rather sparse. Consequently, there are still many areas open for further research. The discussed methodologies for estimating the diluted treatment effect each depend on assumptions that may be too restrictive in certain applications. An additional challenge concerns bias of standard ATE estimators induced by triggering. \citet{chen2018automatic} identify a special type of bias that occurs when a user triggers on day $d$, but not day $d+1$. Other types of bias, as well as the questions of how to define the randomization unit (user, session, or webpage) and how and when to aggregate data into sessions, remain open for exploration. Recent work in \citet{deng2023zero} also suggests that exploring noncompliance and other similar concepts from the causal inference literature (such as principle stratification) with respect to triggering may be an area for future development.


\section{Heterogeneous Treatment Effects}
\label{sec:hetero_effects}

\noindent \textbf{Motivating Example:} \textit{Suppose an online ad provider wishes to determine the impact of changing from static textual ads to short video ads on website traffic. For the treatment group, website traffic appears to have increased uniformly except among Safari users. Consequently, the ad team wishes to estimate the treatment effect at the browser level. Likewise, after observing an improvement in user engagement metrics, the ad team may want to perform a post-hoc analysis to determine if this increase is roughly the same for all users or perhaps concentrated within certain user segments (such as those defined by market/country, user activity level, device/platform type, and time).}

Treatment effects on subgroups that differ from the population-level ATE are known as \textit{heterogeneous treatment effects (HTE)} and are commonly of interest in OCEs. Identifying and interpreting such heterogeneity is vitally important for business applications. For example, practitioners may be interested in estimating the treatment effect for different devices or browser types, or for users of different ages, or users living in different parts of the world. 
Identifying and estimating HTEs is also of concern for those wishing to develop personalized experiences, or to detect bugs, or interactions with other experiments. Three key challenges are associated with estimating HTEs: (1) small treatment effects (see Section \ref{sec:sensitivity_and_small_effects}) often make online studies under-powered, resulting in high false negative rates for subgroup effects; (2) testing for HTEs tends to risk inflated Type I error rates due to multiple comparisons; (3) in cases where users are not randomized to the subgroups under comparison, the usual tension between correlation and causation manifests. Below we review existing methodologies that are commonly used in the context of OCEs to address this problem.

Heterogeneous treatment effects have a rich history in statistical theory and application \citep{robinson_1988, athey_heteroPartitioning_2016, wagner_heteroRF_2018, yingqi_2012, tran2019learning, imai_hte_2013}. In this review, we focus on the intersection of this literature and OCEs. Assume each unit $i$ has a pair of potential outcomes $\big\{Y_i(1), Y_i(0)\big\}$ and a vector of pre-treatment covariates ${X}_i$, with $e(x) = Pr(W_i = 1|X_i = x)$ being the probability that a user is treated given a particular value of the covariate. For randomized studies where causality may be inferred, $e(X_i)$ is known and technically independent of ${X}_i$; however, when HTE analysis is under an observational setting, $e(X_i)$ is typically unknown. Most of the literature regarding HTEs employs the following assumptions: (1) SUTVA and (2) \textit{unconfoundedness}, meaning that the response is independent of the treatment assignment $W_i$ conditional on the covariate, $\big\{Y_i(1), Y_i(0)\big\} \ \indep \ W_i | X_i$. The main goal is to estimate the \textit{conditional average treatment effect (CATE)}, $\tau(x) = \E[Y_i(1) - Y_i(0) | X_i = x ]$. Another key challenge is to detect exactly for which specific levels of the covariate $\tau(x)$ differs from $\tau$ and, given several covariates, identifying which $X$'s are the source of heterogeneity.

Interpretation is crucial in the online industry, thus a popular approach is to assume a linear mapping from $Y_i$ to $\big(W_i, X_i\big)$ from which main and interaction effects may be estimated. Unsurprisingly, the relationship between $Y_i$ and $X_i$ is often highly complex, thus a common method is to use the semi-parametric model from \citet{robinson_1988}, $Y_i = \tau(X_i)W_i + g(X_i) + \eps_i$, which makes no assumptions about the forms of $\tau(X_i)$ and $g(X_i)$. Under unconfoundedness, one may write $Y_i - m(X_i) = \tau(X_i)(W_i - e(X_i)) + \eps_i$, where $m(X_i) = \E[Y_i|X_i]$ and $e(X_i)$ are unknown. The $\ell_2$ loss function is used to estimate the heterogeneous treatment effects, resulting in the estimate $$\hat{\tau}(X) = \argmin_{\tau'}~\frac{1}{n}\sum_{i=1}^n \left\{Y_i - m(X_i) - \tau'(X_i)\left[W_i - e(X_i)\right]\right\}^2.$$ Thus HTE estimation is a ripe target for machine learning methods. Researchers have approached this problem using a technique called ``Double Machine Learning" (DML) \citep{chernozhukov2017double,}. Briefly, this technique models $m(X)$ and $e(X)$ as nuisance parameters, estimating them with nonparametric regression on a hold-out sample set. The CATE may then be estimated on the remaining sample set using a variety of machine learning methodologies. \citet{chernozhukov2017double} demonstrate that the above squared error loss is Neyman orthogonal to $m(X)$ and $e(X)$, which, along with sample splitting, ensures unbiasedness of $\hat{\tau}(X)$ and enforces parsimonious modeling of $Y$ with respect to the nuisance parameters. \citet{syrgkanis2019machine} extend DML for estimating heterogeneous treatment effects when the covariates are hidden. Such situations arise in online experiments when users do not comply with a treatment due to unobserved factors. By modeling $Y$ with instrumental variables, \citet{syrgkanis2019machine} estimate the HTEs using a doubly-robust, fully convex loss function that is minimized with an algorithm that builds on the DML technique. To avoid the challenge of directly estimating the HTE, \citet{peysakhovich2016combining} utilize historic, user-level data to learn individual effect estimates conditional on the covariates that correlate with the true treatment effect. Practitioners at Netflix also used DML to understand the localized impact on viewership from subbed and dubbed movies \citep{netflix_dml}.

Other popular machine learning approaches for estimating $\tau(X)$ are regression trees and random forests. Following the DML approach, for instance, one may use trees to identify meaningful segments of a continuous or categorical variable, and then model $\tau(X)$ with partially linear regression. In an adaptation of the classical CART algorithm, \citet{athey_heteroPartitioning_2016} build modified regression trees to partition the data into subgroups corresponding to different magnitudes of the treatment effect, thus each terminal leaf produces an estimate for $\tau(x)$, rather than the traditional estimate of $\E[Y|X_i=x]$. To correct for over-fitting, an additional split of the training data into non-overlapping sub-partitions for each leaf is used. Naturally, this method can be extended to random forests to create a causal forest for estimating the HTE \citep{wagner_heteroRF_2018}. While casual trees and forests do not require linearity of the treatment effects, and perhaps are conceptually more intuitive than DML, they are somewhat lacking in terms of interpretability compared to the effect estimates from DML and other similar methods. A further disadvantage is that the additional training split reduces the sample size for an application that may already suffer low power.

In addition to estimating $\tau(X)$, identifying which covariates or levels of covariates contribute to treatment heterogeneity is of great practical concern \citep{sepehri2020interpretable,mcfowland2021prescriptive}. Obtaining a parsimonious model of $Y$ is critical in such situations, as there are typically a large number of covariates from which to choose and strong statistical significance is required for detecting HTEs. This challenge is simultaneously a variable selection and multiple testing problem. \citet{xie_false_2018} assume an experimental design setup where $e(X) = Pr(W_i = 1)$, using this value to transform $Y_i$ into $Y_i^*$ such that $\E[Y_i^* | X] = \tau(X)$, which is estimated with the standard difference-of-means $\hat{\tau}$. Using $Y^* - \hat{\tau}$ as the response variable, the authors perform Lasso regression in conjunction with the ``knockoff" variable selection defined by \citet{barber2015controlling} to select heterogeneous covariates while controlling false discovery rates. They also demonstrate how to use the Benjamini-Hochberg correction to identify levels within these covariates where HTEs occur. \citet{deng2016concise} also use variable selection when clusters of covariates are of interest, such as device grouped by brand name. They employ a linear model with first order effect and second order interaction terms and enforce sparsity using total variation regularization, a technique similar to Fused Lasso \citep{petersen2016fused}. 

Given the wide array of scenarios under which HTEs occur in online experiments, there are still many situations where the methods discussed above may not be appropriate. Much of the literature in this review make strong model assumptions that are difficult to verify in practice. For example, the SUTVA requirement that units in a given variant receive the same level of treatment may not be satisfied if users have different levels of engagement \citep{imbens2015causal}; highly engaged users will typically experience a higher ``dose'' of their treatment (relative to light users) due to repeated exposure over the experimental period. As such, the vast literature on dose-response studies \citep{ruberg1995doseI,ruberg1995doseII} seems pertinent here. Additionally, the low power due to small effect sizes makes multiple testing quite challenging. Simulations regarding the approach for controlling FDR in \citet{xie_false_2018} showed that the knockoff method may be too conservative when faced with small effect sizes, and \citet{deng2016concise} reported difficulties regarding high false positive rates. For more open challenges regarding HTE estimation, we encourage the interested reader to consult \citet{gupta_summit_2019, kohavi_trustworthy_2020, bojinov2022online}.


\section{Long-Term Effects}
\label{sec:long_term_effects}

\noindent \textbf{Motivating Example:}
\textit{At Bing, researchers hypothesised that generating large numbers of advertisements should have a positive effect on revenue, but may hurt user engagement in the long-term. To test this, the researchers exposed users to varying ad loads, noting a significant difference in engagement metrics for users exposed to a high ad load versus a low one. It was proposed that one may estimate the long-term effect by performing a post-hoc analysis some time after the experiment. Unfortunately, the post-hoc differences between high-load and low-load users could not be solely attributed to treatment assignment -- many users quickly abandoned Bing as a result of too many ads, biasing results towards the users who remained \citep{dmitriev2016pitfalls}.}

Practitioners are often interested in understanding the treatment effect not just during the experiment, but months, even years after the experiment concludes. In many online experiments, the short-term treatment effect observed during and immediately after the experiment is not necessarily the same as the long-term effect. For instance, click-bait advertising has a positive short-term effect on click-through-rates, but a negative long-term effect on user retention and revenue \citep{kohavi_trustworthy_2012}. More generally, novelty and primacy effects are of concern. A novelty effect exists when a novel change is initially intriguing, leading to increased engagement, but that diminishes over time. A primacy effect on the other hand exists when the initial reaction to a change is not positive, but over time as users get used to the change their engagement increases \citep{mcfarland_experiment_2012,sadeghi2022novelty}. In both cases, the nature of the treatment effect may change over time as users learn. The treatment effect itself may also change dynamically over time independent of user learning. As such, the ATE $\tau$ should not be viewed as a constant with respect to time ($t$), it should more appropriately be regarded as a function of it: $\tau(t)$.

Current OCE literature regarding long-term effect estimation is highly context-specific. At the time of writing this review, it is difficult to pinpoint a single statistical lineage of methodologies for this area (unlike with heterogeneous treatment effects, for example). We begin by introducing several distinct approaches that draw from a variety of statistical fields, and then finish with discussion of one area in particular that shows promise in providing a statistical framework for modeling and estimating long-term effects in online settings. For more industry-specific examples of the challenges concerning long-term effects, see \citet{gupta_summit_2019,bojinov2022online}.

A straightforward way to assess long-term effects is to simply run the experiment longer and ensure that the appropriate metrics for capturing long-term behavior are observed. However, much of the literature written by practitioners of OCEs has been devoted to describing the pitfalls associated with running long-term controlled experiments specifically for estimating long-term effects \citep{kohavi_controlled_2009, kohavi_trustworthy_2012, dmitriev2016pitfalls, gupta_summit_2019, kohavi_trustworthy_2020}. Besides increased cost, several other external factors often make long-term experiments unappealing. For instance, when browser cookies are used to identify users, long-term experiments risk losing upwards of 75\% of users as a result of cookie churn and are rendered invalid as a result \citep{dmitriev2016pitfalls}. These users may also re-enter the experiment unbeknownst to the experimenters and receive both the treatment and control experiences. This type of contamination can also happen if users access the product or service on multiple devices, a problem that becomes more likely as the experiment's duration increases. Additionally, the longer the experiment the more likely it is that multiple users (e.g., family members) will use the same device, obfuscating results. As such, in this section, we focus on techniques for estimating long-term treatment effects alternative to increasing experiment length. 

Several approaches for estimating long-term effects intersect with other areas discussed in this review. In \citet{wang2019heavy}, long-term effects are characterized as a form of bias due to heterogeneous treatment effects (Section \ref{sec:hetero_effects}). In this context, long-term effects manifest because heavy-users (frequent users of the product) tend to be included in experiments at higher rates than light-users, biasing the ATE particularly in the short-term. Here, the treatment effect is presumed to be different depending on whether user $i$ is a heavy- or light-user. Under SUTVA and an assumed independence of outcomes from treatment assignment, the authors derive bias due to heavy-users in closed form, proposing a bias-adjusted jackknife estimator for the overall ATE. For a two-sided market where each experimental unit has a treatment history up to time $t$, \citet{shi2020reinforcement} leverage sequential testing (Section \ref{sec:optstop}) and reinforcement learning to test for long-term treatment effects. Using data from a ride-sharing company, they demonstrate how their derived test statistic is able to detect long-term effects where regular two-sample t-tests fail. While the solutions from \citet{wang2019heavy} and \citet{shi2020reinforcement} are effective, they only target specific types of long-term effects, which limits their potential generalizability to other settings. 

Another common solution is to define and measure short-term \textit{driver metrics} that are causally linked to the long-term effect \citep{kohavi_trustworthy_2020}. Driver metrics allow practitioners to focus experiments on short-term goals while still taking into account the long-term effects (see \citet{netflix_proxy_metrics} for anecdotal examples). In \citet{hassan2013beyond}, the authors define heuristics for modeling implicit indicators of customer satisfaction, noting that using query-based models instead of click-based models tend to serve as better proxies. \citet{hohnhold_focusing_2015} define models for how users ``learn" to search or click for a product as a result of being exposed to a treatment, such as change in number of ads shown, using ``learned click-through-rates" as a driver metric for estimating long-term effect on revenue. To estimate the effect on long-term revenue using short-term effects due to treatment, the authors model this as a linear function of short-term revenue and the estimated learned click-through-rates. The model has been successfully deployed by Google and is widely cited in the OCE literature \citep{kohavi_trustworthy_2020, gupta_summit_2019, deng_trustworthy_2017, wang2019heavy}. A recent paper by \citet{sadeghi2022novelty} proposes an observational approach based on difference-in-differences to estimate user learning and hence the long-term treatment effect in contexts where novelty and primacy effects exist.

Methodology in this area tends to resemble recent works in the causal inference literature that also aim to address this challenge by combining short-term experimental data with long-term observational data. In Section SM2 of the Supplementary Material we elaborate on the use of \textit{surrogate outcomes} \citep{prentice1989surrogate, begg2000use, frangakis2002principal, ensor2016statistical} in this context.


\section{Optional Stopping}
\label{sec:optstop}

\noindent \textbf{Motivating Example:} \textit{Suppose an online streaming service is altering a certain feature that positively correlates with subscription renewals. While an improvement to this feature could increase the rate of subscription renewals, a harmful change may have the opposite effect. It is in the service's best interest to quickly abandon harmful or poorly performing variants, and identify those that perform well. Methods that support early termination without compromising overall statistical validity are desirable. A notable practice within this context is to ``ramp up” the experiment by slowly exposing an increasing percentage of users to the treatment \citep{xu_sqr_2018}, and to integrate OCEs into a phased software deployment via ``controlled rollout'' \citep{xia2019safe}.}

Most OCEs are run in real-time, and it is not uncommon for estimates and confidence intervals associated with $\tau$, and p-values associated with $H_0:\tau=0$, to be updated in near-real-time as the data are collected. Although a fixed horizon is typically determined based on development cycles (typically two weeks) and minimal sample size requirements (determined via power arguments), the near-real-time availability of results encourages a phenomenon known colloquially as ``peeking," whereby p-values are monitored continuously and the experiment is stopped as soon as a significant p-value is observed. While it is well known that this practice seriously inflates false positive rates \citep{johari2022always, kohaviIntuitionBusters}, there are nevertheless situations where having a mechanism for \textit{optional stopping} is desirable. For example, it is extremely important to quickly detect and abort treatments that are negatively impacting the user experience \citep{lindon2022rapid}. Thus, in situations like this, a methodology that permits near-real-time decision-making without inflating Type I error rates is invaluable. Unsurprisingly, design and analysis methods from the \textit{sequential testing} body of literature are of relevance here. Below we describe the development and application of methods in this area to the OCE context.

As per the field of sequential testing, interest lies in assessing $H_0:\tau = 0$ using sample size-dependent decision rules. Within this class of methods, Type I error is controlled at each current sample size $n$, which avoids the inflated risk of Type I error that is associated with preemptively stopping an experiment when the current p-value is statistically significant by chance. Such methods improve testing efficiency due to the lower sample size a sequential test will terminate at, on average, regardless of where the true treatment effect might lie. However, there is no free lunch. Existing methodology is not well-suited for all OCE applications, such as monitoring multiple metrics (e.g., the OEC and guardrails). Additionally, the reduced sample sizes guarantee under-powered HTE inference across user segments. Nevertheless, sequential testing methods have appeared in numerous OCE applications and studies \citep{kohavi_online_2013, johari2022always, kharitonov2015sequential, deng2016continuous, yu2020new, shi2020reinforcement, abhishek_2017_nonparametric_sequential, ju_2019_sequential_ab_test,spotify2023sequential,booking2023sequential}. The following section broadly introduces the method of sequential testing as it pertains to ongoing evaluation of the treatment effect(s) of interest in OCEs. 

The majority of the OCE literature in sequential testing builds on the classic \textit{sequential probability ratio test} (SPRT) developed by \citet{wald_1945_sequential}. Define constants $0 < B < A$ where $B = \beta/(1-\alpha)$ and $A = (1-\beta)/\alpha$, and a simple hypothesis test $H_0: \theta = \theta_0$ versus $H_1: \theta = \theta_1$. The SPRT method proceeds as follows. For current sample size $n$, compute the likelihood ratio test statistic $$\Lambda_n = \prod_{i=1}^n \frac{f(y_i|\theta_1)}{f(y_i|\theta_0)},$$ where $y_i$ are observations of i.i.d data $\{Y_i\}_{i=1}^n \sim f(\cdot|\theta)$. The rejection region divides the sample space into three mutually exclusive decision rules: (1) if $\Lambda_n > A$, reject $H_0$ and stop the test. (2) If $\Lambda_n < B$, fail to reject $H_0$ and stop the test. (3) If $B < \Lambda_n < A$, obtain another observation $Y_{n+1}$ and compute $\Lambda_{n+1}$. Although it seems like testing in this manner would permit the possibility of never drawing a conclusion about $H_0$ (i.e., $n \rightarrow \infty$), \citet{wald1947sequential} proved that the SPRT will eventually terminate for finite $n$. SPRT does not require specifying $n$ in advance, and requires on average about half the number of observations required for a uniformly most powerful Neyman-Pearson test for the same level of power \citep{wald_1945_sequential}.

The first and perhaps most widely-known application of sequential testing in OCEs is a modified version of SPRT called the \textit{mixture sequential probability ratio test}, or mSPRT \citep{johari2022always, pramanik2021modified}. The mSPRT allows for a simple null hypothesis versus a composite alternative hypothesis $H_1: \theta \neq \theta_0$ by assuming a mixture distribution $H$ with density $h(\cdot)$ defined over the parameter space of all possible $\theta$. The test statistic is therefore a mixture of the likelihood ratios, $$\Lambda_n^H = \int \prod_{i=1}^n \frac{f(y_i|\theta)}{f(y_i|\theta_0)}h(\theta)d\theta.$$ The procedure rejects $H_0$ and ends if $\Lambda_n^H \geq \alpha^{-1}$. \citet{johari2022always} use the mSPRT to define ``always valid p-values", which are computed iteratively such that $p_0 = 1; \ p_n = min\{p_{n-1}, (\Lambda_n^H)^{-1}\}$. Thus, practitioners may stop an experiment at any time while still controlling Type I error. The always valid p-values and their confidence interval counterparts are currently deployed by Optimizely, a widely-used third-party vendor for OCEs \citep{optimizely_alwaysvalid}. With the vast quantity of experiments that this company has facilitated, they are able to leverage prior data for estimating the mixture distribution $H$. However, \citet{johari2022always} derive their optimality conditions for mSPRT only for data that comes from the exponential family of distributions, which does not include distributions for the ratio metrics that are popular in industry. Another limitation lies in how the likelihood ratios for a two-sample hypothesis test are defined. While the authors assume a standard independent, two-sample stream of data, they impose an additional restriction by arbitrarily pairing observations, which suggests a matched pairs design. This allows for defining a tractable $f(y_i | \theta)$, but there is no practical reason for observations to be paired, and no practical guidance given for how to perform the pairing. Additionally, the assumption that observations arise independently may be violated when a unit generates multiple observations as they interact repeatedly with the experiment over time. Moreover, although the Type I error rate is satisfactorily controlled (when all assumption are met), unbiased estimation of the treatment effect is still a concern. Methodology that relaxes these assumptions and yields unbiased estimates is therefore valuable.

The well-publicized usage of mSPRT has inspired several related works in the literature. \citet{abhishek_2017_nonparametric_sequential} account for the situation where $f(\cdot|\theta)$ is unknown by creating a bootstrap algorithm to approximate $\Lambda_n^H$. While the algorithm also requires a prior distribution to approximate $H$, this method still allows practitioners to use mSPRT for commonly used online metrics that are otherwise difficult to model.  The work in \citet{lindon_anytime_2020} extends mSPRT to multinomial count data, which includes an application for conducting SRM tests sequentially, in near-real-time. \citet{yu2020new} also extend mSPRT to the multiple testing scenario to test for heterogeneous treatment effects, using always valid p-values to allow for continuous monitoring. In a well-received paper, \citet{xu_sqr_2018} use a technique similar to mSPRT, called a \textit{generalized sequential probability ratio test} (GSPRT), to determine the risk of exposing more users to a new variant. Briefly, the GSPRT uses the supremums of the likelihoods in $\Lambda_n$ and can be shown to require smaller sample sizes on average than mSPRT \citep{chan_gsprt_2005}. \citet{xu_sqr_2018} use a prior-weighted GSPRT to provide a rigorous statistical framework dubbed ``speed, quality, and risk" (SQR) for the practice of ramping up, gradually introducing users to a new variant in order to mitigate the fallout associated with exposing them to potentially negative variants (for a high-level discussion of SQR, see Chapter 15 of \citet{kohavi_trustworthy_2020}). An alternative to frequentist sequential testing is also explored by \citet{deng2016continuous}, where the authors use Bayesian hypothesis testing as the foundation. Bayesian methods for OCEs are briefly discussed in Section SM4 of the Supplementary Material. 

Finally, we acknowledge that the sequential testing methods discussed here are fully sequential. However, group sequential methods commonly used in adaptive clinical trials \citep{pocock1977group,o1979multiple,lan1983, robertson2023point} are rapidly gaining in popularity in the context of OCEs.  \citet{georgi2022gst, booking2023sequential, spotify2023sequential} describe the use of these methods in this context, and the value they provide with respect to the speed of decision making (i.e., increased power) and false positive control. The adaptation of such methods to tailor them for use with OCEs seems like a nascent though fruitful line of research.


\section{Interference}
\label{interference}

\noindent \textbf{Motivating Example 1:} \textit{Suppose LinkedIn plans to test the impact of a new feature for their messaging service, with the objective to increase total messages sent. Using balanced randomization, given that user $i$ is exposed to the new feature, there is approximately a 50\% chance said user's friend $j$ is randomized to the old service. Under this scenario and if the new feature indeed increases messages sent, it is likely that friend $j$ will also send more messages in response to $i$, despite $j$ belonging to the old service. Thus, the overall impact of the new messaging feature on total messages sent is confounded by the network interference between treatment and control groups, biasing standard estimators for the ATE \citep{linked_network_abtest_example}.}

\noindent \textbf{Motivating Example 2:} \textit{Suppose Lyft is experimenting with a new version of the pricing algorithm that results in treated passengers booking more rides. However, since the number of available drivers is finite, increased ride bookings in the treatment group necessarily reduces the supply of drivers and hence the number of possible rides for the control group \citep{lyft2016interference}. This in turn biases naive treatment effect estimates that compare ride bookings in the two groups.}

Recall that SUTVA requires that the potential outcome $Y_i(W_i)$ for unit $i$ remain the same regardless of the treatment assignments and outcomes of the other experimental units. However, in certain OCE applications (e.g., social networks and online marketplaces), SUTVA may be violated when the treatments \textit{interfere} with each other. Such a SUTVA violation is referred to as \textit{interference, spillover,} or \textit{leakage}. Interference was illustrated in both motivating examples above; in each case a unit's outcome depended not only on its own treatment assignment, but also on the treatment assignments and outcomes of other units in the experiment. The two examples typify two different forms of interference: the first, \textit{network interference}, arises when the units are connected to one another through a network, such as a social network like LinkedIn \citep{saint2019using} or Facebook \citep{eckles_design_2014}. The second form of interference, \textit{marketplace interference}, arises when units compete for shared resources in two-sided marketplaces such as Lyft \citep{lyft2016interference}, Ebay \citep{blake2014marketplace}, or Airbnb \citep{holtz2020reducing}, and three-sided marketplaces like DoorDash \citep{tangcontrol}. Note that \citet{kohavi_trustworthy_2020} defines these interference mechanisms as resulting (respectively) from ``direct'' and ``indirect'' connections among the units. \citet{bojinov2022online} alternatively define \textit{partial} and \textit{arbitrary} interference in addition to marketplace interference. Whether the interference mechanism is partial or arbitrary depends on whether a unit is influenced by some or all of the other units in the experiment. \citet{kohavi_trustworthy_2020} describe another type of interference whereby a malfunctioning treatment causes a crash that impacts both treatment and control users. 

As described in Section \ref{sec:notation}, estimates of the average treatment effect seek to quantify the difference in outcomes when all units are treated versus when all units are controlled. In standard settings, a subset of units randomized to treatment and another subset randomized to control serve as an adequate proxy for the unobserved counterfactuals. However, when the treatment and control groups interfere with each other, traditional randomization no longer adequately approximates the counterfactuals and hence standard difference-in-means estimators are no longer unbiased. A rich literature has recently been developed that carefully considers both the design and analysis of OCEs in the presence of interference. Of particular relevance are experimental designs that reduce the amount of interference, and analysis methods that provide unbiased treatment effect estimates in the presence of interference. We provide a brief overview of that literature below. As will become apparent, many ideas from the clinical RCT literature are relevant here (e.g., cluster randomized trials and crossover designs).

In network A/B tests, cluster-based randomization methods have gained widespread attention \citep{karrer2021network, eckles_design_2014, gui_network_2015, saveski_detecting_2017, sangho_yoon_designing_2018, zhou_cluster-adaptive_2020, ugander_graph_2013}. Such methods first involve partitioning the network into mostly disjoint clusters, commonly via community detection algorithms. Such a partitioning yields groups of nodes with much greater intra-cluster connectivity than inter-cluster connectivity. Randomization is then performed at the cluster level, where all units in a cluster receive the same treatment assignment. In doing so, units will (for the most part) have the same treatment assignment as the units nearest--and hence most likely to influence--them. This limits the opportunity for interference and therefore better mimics the all-treated and all-controlled counterfactuals. However, with the clusters being the randomization unit (instead of the individual users), the effective sample size and hence power is dramatically reduced. \citet{saint2019using} therefore propose the use of many, smaller \textit{ego-clusters}. These clusters are defined by a single user (the ego) and a subset of its direct neighbours (the alters). Ego-cluster-based randomization does not pay as significant a penalty in terms of power, since there are many more ego-clusters than in a traditional cluster-based design. Moreover, other treatment assignment schemes in which the ego is treated differently from the alters may be used to estimate the network interference. In Section SM3 of the Supplementary Material we elaborate on these and other methods developed for the design and analysis of network A/B tests. 

The spirit of cluster-based randomization--that is, to treat units likely to influence each other in the same way--is at the heart of methodologies intended to mitigate other forms of interference as well. For instance, in settings with marketplace interference, \textit{switchback} experiments are commonly used to sequentially alternate units between treatment and control over time \citep{bojinov2022design}. In doing so, at any given time period all users have the same treatment assignment and therefore cannot exert influence on each other. And then through repeated exposure to treatment and control over time, the ATE can be estimated. Such experiments suffer from temporal carryovers, but these can be mitigated with ``burn-in'' periods analogous to washout periods in clinical trials \citep{hu2022switchback}. Optimal design strategies have also been developed to address carryover effects \citep{bojinov2022design}. Even still, like cluster-based randomization in the network setting, switchback experiments suffer from decreased power via a reduced effective sample size. Recent work by \citet{ni2023design} explores the use of spacial clustering and temporal balance to overcome this problem. 

In marketplaces based on auctions (e.g., eBay auctions \citep{blake2014marketplace} or advertising auctions \citep{liu2021trustworthy}), interference due to shared resources is also a problem. A treatment that encourages higher bidding will lead to treated users winning more auctions and control users necessarily losing them. This leads to a ``cannibalization bias'' whereby the margin by which the treatment looks better than the control is exaggerated, because when the treatment wins, the control must lose. Switchback experiments have been proposed as a means to mitigate such bias, but their limitations (described above) have led to the development of more tailored experimental designs for online auctions. For example, \citet{liu2021trustworthy} propose \textit{budget-split} designs that eliminate the opportunity for cannibalization bias by splitting the available resources (i.e., the budget) equally and independently between the treatment and control groups. With the resources no longer shared, there exists no imbalanced competition for them.

The works described above seek to eliminate (or at least minimize) interference through the experiment's design so that the traditional difference-in-means estimator yields unbiased estimates of the ATE. However, an alternative paradigm exists in which interest lies in de-biasing through the analysis of the experiment by \textit{modeling} the interference rather than eliminating it. For instance, \citet{bui2023general} develop a class of general additive network effect models that facilitate unbiased ATE estimation while flexibly modeling network influence. Many other such network modeling approaches exist. See, for example, \citet{parker2017optimal,basse2018model,pokhilko2019d,koutra2021optimal,zhang2022locally} and Section SM3 of the Supplementary Material for a deeper discussion of such methods. Likewise, \citet{johari2022experimental} and \citet{li2022interference} develop stochastic market models to capture interference dynamics in two-sided marketplaces. Such methods, if the interference is accurately modeled, enjoy the benefit of increased power by permitting randomization at the user-level. However, accurately modeling interference is non-trivial.



\section{Conclusion}
\label{sec:the_end}

The value of experimentation and the accompanying philosophy of trial and error has been observed in many facets of society \citep{basic2019uncontrolled}, and its positive impacts in the realm of business in particular are remarkable \citep{koning2022experimentation}. Online controlled experiments are vital tools utilized hundreds of times a day by companies whose products touch the lives of billions \citep{kohavi_online_2013, xu_infrastructure_2015, HowGoogl12:online}. As many vital societal functions shift online at an unprecedented rate, online experimentation has already found applications outside the mainstream spheres of technology and e-commerce. OCEs have been used to optimize political advertisements and increase user engagement with campaign platforms during the Obama and Trump elections \citep{christian_b_2012, bump_2019}. Decision-making tools streamlined by OCEs help clinicians make safer, more cost effective decisions regarding patient care \citep{austrian2021applying}. OCEs have also been deployed to identify the psychological impacts of social media on younger demographics \citep{isaac_2021}. Along with the significant growth and popularity of careers under the evolving ``data science" profession, online experimentation is almost certainly going to become a common tool for online businesses of all sizes \citep{Forbes_data_analytics}. Given the breadth and depth of OCE applications, we believe that solving the research challenges presented in this review will improve the quality of data-driven decision making in online businesses across the applied domain. 

We conclude this literature review with a call to action for greater collaboration between industry and academic statisticians to address the research challenges presented by online experimentation. While this paper may be one of the first to provide a cohesive review of the OCE statistics literature, the need for increased cooperation between industry and academia already has been explicitly stated by experts at thirteen leading organizations that run online experiments \citep{gupta_summit_2019}. Collaborative partnerships between academia and industry do exist in this space (see, e.g., \citet{waudby2021time,ham2022design} respectively for partnerships between Carnegie Mellon and Adobe and Harvard and Netflix on the problem of sequential experimentation). However, the academic statistics community in general seems to lack familiarity with--and access to--this research area. The purpose of this review, therefore, was to introduce academicians to the context and goals of online experimentation, as well as to provide examples and broad, technical discussion of the statistical methodologies regarding sensitivity, effect size, heterogeneity, long-term effects, optional stopping, and interference. In the absence of direct collaboration with industry, it is difficult to develop and test novel methodology without access to data.  While \emph{some} open access repositories exists (see, e.g., \citet{liu2021datasets,matias2021upworthy}), the proprietary nature of these experiments makes open-access data-sharing uncommon. This is admittedly a challenge and a remaining open problem for research in this space.

\section*{Acknowledgements}

The authors thank Art Owen, Georgi Georgiev, and Somit Gupta for helpful comments on an earlier draft of this paper.


\section*{\centering SUPPLEMENTAL MATERIAL}\label{sec:sm}
\setcounter{section}{0}
\renewcommand{\thesection}{SM\arabic{section}}

\section{Stratified Sampling and CUPED}\label{sec:stratSM}

Assume there exist $K$ strata dividing the population $\Omega$, where every stratum has mean and variance $(\mu_k, \ \sigma_k^2)$, and each unit $i$ falls into the $k^\text{th}$ strata with unknown probability $w_k$ such that $\sum_{k=1}^Kw_k = 1$. With data obtained via stratified sampling, it is well-known that one may construct an unbiased, weighted estimator of $\tau$ that has smaller variance than the standard difference-of-means estimator, presuming one has correctly estimated $w_k$ and identified stratum that are correlated with $Y$ \citep{acharya2013sampling}. As noted in \citet{deng_improving_2013} and \citet{xie_improving_2016}, many organizations have access to large amounts of data, which can simplify the process of identifying meaningful strata. However, estimating $w_k$ is not straightforward, and the real-time nature of online experiments as well as the physical infrastructure of experimentation platforms also hinder accurate implementation of stratified random sampling. The primary challenge is to maintain equal representation of the strata while users are randomized to treatment and control. \citet{xie_improving_2016} propose a novel stratified sampling technique that involves defining one queue $q$ for each strata $k$. Each $q$ consists of multiple segments of fixed length. Depending on their strata, users are first assigned to a slot within a segment, then treatments are randomized within each segment. Consequently, balanced allocation is only guaranteed within a segment. Moreover, if multiple machines each have their own $q$ for strata $k$, as is the case in many large experimentation platforms, balanced randomization is even more difficult to achieve. \citet{deng_improving_2013} show that CUPED is equivalent to stratified random sampling when the control variate is categorical, and is considered a post-experiment workaround for the practical difficulties of implementing stratified sampling in real-time. \citet{xie_improving_2016} compare their stratified sampling technique to CUPED, finding that CUPED prevails in terms of variance reduction. Practitioners continue to be interested in methods for stratified sampling with the aim of variance reduction, as well as identifying such strata in order to detect bugs or potential areas for targeted optimization.

\section{Surrogate Outcomes for Long-term Treatment Effect Estimation} \label{sec:lteSM}

This literature generally begins with the following. Assume a potential outcomes setup with two samples, $n_E$ (experimental) and $n_O$ (observational), with binary indicator $G_i \in \{E, O\}$. The tuple $(W_i, \ S_i, \ X_i)$ is observed in the experimental group and $(Y_i, \ S_i, \ X_i)$ in the observational group, where $S_i$ is an intermediate short-term outcome and $X_i$ is a pre-treatment covariate ($W_i$ may also be included in the observational group, see \citet{athey2020combine, imbens2022longterm}). The goal is to estimate the average treatment effect of $W_i$ on $Y_i$, which is nontrivial since $Y_i$ is not observed in the experimental sample. The origins of this framework can be traced back to statistical literature regarding \textit{surrogate outcomes}, used largely in biostatistics and econometrics \citep{prentice1989surrogate, begg2000use, frangakis2002principal, ensor2016statistical}. The work by \citet{athey2019surrogate} is one of the first papers that uses this framework for long-term effect estimation cited within the OCE community. The authors derive estimators of $\tau$ using $S_i$ as driver metrics and assume $W_i$ is not observable in $O$. They employ the ``surrogate criterion", which requires that $Y_i$ be independent of $W_i$ given the short-term outcomes. It is straightforward to see that the approach in \citet{hohnhold_focusing_2015} is a special case of this approach, where $S_i$ is comprised of the learned click-through-rates and short-term revenue, $Y_i$ is long-term revenue, and the necessary conditions for estimating $\tau$ are unverified but implicitly assumed.

In practice, the surrogate criterion is notoriously tricky to satisfy. \citet{athey2020estimating} relax this assumption by only requiring that $Y_i$ is independent of $W_i$ conditional on a \textit{set} of surrogates, rather than on each individual surrogate. In perhaps one of the earliest publications using statistical surrogacy to estimate long-term effects specifically in OCEs, \citet{cheng2020long} show that one can relax the surrogacy assumption by extending this framework to incorporate sequential testing. There is also evidence that some tech companies such as Facebook have used statistical surrogacy \citep{gupta_summit_2019}, although it appears that too many surrogates may severely hamper interpretability. Recent work has shifted away from the surrogate criterion. \citet{athey2020combine} let $W_i$ be seen in the observational sample and estimate the treatment effect on $S_i$ in both samples, using the difference to adjust the ATE estimates. \citet{imbens2022longterm} consider a similar context and demonstrate how to account for unmeasured confounding variables that impact treatment, short-term, and long-term outcomes.  \citet{van2023estimating} point out that surrogate methods that assume there are no unobserved confounders in the observational data may not be a practically useful. As an alternative they propose an instrumental variables approach to estimate a long-term effect by combining regression residuals with short-term experimental outcomes. Further exploration of combining short-term experimental data with observational data to estimate long-term effects may show promise with respect to OCE applications.

\section{Network A/B Tests} \label{sec:netSM}

OCEs where the experimental units are subject to network exposure are known as \textit{network A/B tests}, where users and the connections among them are modeled by a network $\mathcal{G}$, with $n \times n$ adjacency matrix $\mathbf{A} = [A_{ij}]$. In most OCE settings, $\mathbf{A}$ is assumed to be fixed and observable, although situations where this is not the case are also considered \citep{egami2017unbiased}. The goal of estimating the ATE remains of primary interest. However, when SUTVA is violated, standard randomization schemes and estimators tend to ignore the network effect, which typically produces biased estimates of $\tau$. Consider the following example: suppose the response $Y_i = \alpha + \beta W_i + \gamma S_i + \eps_i$ is linearly related to the treatment effect $\beta$ and network spillover effect $\gamma$, where $S_i$ is the proportion of $i$'s neighbors that received treatment. The ATE is therefore $\beta + \gamma$, since $\E[S_i | W_i = 1] = 1$ and $\E[S_i | W_i = 0] = 0$. Under the usual balanced randomization, however, $\E[S_i] = 0.5$ for both treatment and control groups, thus the expected value of the usual difference of means estimator $\hat{\tau}$ is $\beta$, which has a bias of $\gamma$. Generally, the exact form of the ATE depends on the assumed structure of $\mathcal{G}$  and definition of $S_i$; similarly for the form and bias of $\hat{\tau}$. Thus, there are two major problems in network A/B testing that current research aims to address: (1) modeling and estimating the network spillover effect, and (2) optimal treatment allocation for producing unbiased estimates of $\tau$ in the presence of network interference. Reviewing work in these areas is the focus of the following subsection  

A commonly proposed approach for dealing with network effects in OCEs is to randomize treatments with \textit{graph cluster randomization} \citep{karrer2021network, eckles_design_2014, gui_network_2015, saveski_detecting_2017, sangho_yoon_designing_2018, zhou_cluster-adaptive_2020, ugander_graph_2013}. With cluster-based randomization, the network is partitioned into subgroups or \textit{clusters}, such that edge connectivity within clusters is higher than between clusters. Network partitioning, also known as community detection in network science, is a well-researched area, with most OCEs using established graph clustering algorithms as found in \citet{newman_modularity_2006, leskovec_empirical_2010, mucha_community_2010} and \citet{stanley_clustering_2016}. Treatments are then randomized to users at the cluster level with the standard difference of means estimator, a common choice for estimating the ATE. \citet{eckles_design_2014} explore several linear models for relating user response to the network effect, and perform a suite of simulations that show graph cluster randomization reduces bias when compared to naive random allocation. They also provide a theorem that shows the bias from network effects will always be less than or equal to the bias from random allocation, assuming $Y_i = \alpha + \beta W_i + \gamma S_i + \eps_i$. \citet{gui_network_2015} draw from this work, modeling the response as $Y_i = \alpha + \beta W_i + \gamma \sum_{j=1}^n A_{ij}W_j + \eta \sum_{j=1}^n A_{ij}Y_j/d_i$, where $d_i$ is the degree of node $i$, $\gamma$ is the spillover effect, and $\eta$ describes how users tend to exhibit behavior similar to their neighbors'. They showed that with a network sampled such that clusters are ``balanced", where clusters are all equal in size, one can eliminate the bias in $\hat{\tau}$. Their new algorithm for balanced cluster-based randomization was empirically shown to reduce bias, although theoretical justification was not provided. To address the question of how to detect when the spillover effect is present, \citet{saveski_detecting_2017} develop a model-free two stage cluster-randomization design for testing for the presence of SUTVA violations, and \citet{athey2018exact} derive exact p-values for nonsharp null hypotheses of no spillover effects. Recent work by \citet{karrer2021network} utilizes imbalanced clusters with a regression-adjusted estimator, along with a post-analysis framework that is also used to detect network effects. 

While \citet{gui_network_2015} use a common framework for OCE applications, the linear model assumption is known to be quite restrictive, particularly for network applications. \citet{basse_limitations_2018} specifically study the drawbacks of traditional parametric assumptions for modeling network effects. Some practioners instead use \textit{network exposure models} to model the spillover effect \citep{backstrom_network_2011, katzir_framework_2012}. Network exposure models define a set of conditions for each $i$ under which the spillover effects from $i$'s neighbors are the same. For example, the \textit{neighborhood exposure model} from \citet{backstrom_network_2011} and \citet{gui_network_2015} estimates $\tau$ with $\frac{1}{|N^\theta_1|}\sum_{i\in N^\theta_1}Y_i - \frac{1}{|N^\theta_0|}\sum_{i\in N^\theta_0}Y_i$, where $\sigma_i$ is the percent of neighbors of $i$ that received treatment, $N_1 = \{i: W_i = 1, \sigma_i \geq \theta\}$, $N_0 = \{i: W_i = 0, \sigma_i \leq 1 - \theta\}$, and $\theta \in [0,1]$. With network exposure models, one need not make explicit assumptions about how the spillover effect relates to the response, although the corresponding ATE estimators tend to be more complex. \citet{ugander_graph_2013} catalogue the various network exposure models that have been commonly adopted in the literature \citep{eckles_design_2014, gui_network_2015, saveski_detecting_2017}.

While cluster-based randomization approaches are commonly used in practice, the limitations of this method are significant enough that researchers remain interested in alternative approaches. First, because this approach uses clusters as the experimental units and cluster counts typically are far smaller than the total number of users, experiments under this approach tend to lack adequate power. To mitigate this, \citet{saint2019using} propose sampling many ``ego-networks", which are \textit{small} clusters comprised of a central user and a carefully selected subset of their immediate neighbors. Second, the majority of online social networks are highly dense, making it extremely difficult to obtain reasonably isolated clusters that are representative of the true network. \citet{nandy_b_2020} avoid explicit model assumptions by defining $\mathcal{G}$ as a directed network of producers $j$ and consumers $i$. Treatment intervention ($r$) is represented by rewiring edge probabilities by replacing the original $p_{ij}^{base}$ with $p_{ij}^{(r)}$, where $p_{ij} = Pr(A_{ij} = 1)$. \citet{nandy_b_2020} use this setup to frame treatment allocation as an optimization problem, where treatments are randomized such that the effect from network exposure under the new treatment is as small as possible. Their method showed an improvement over cluster-based randomization in terms of bias of the ATE, particularly for highly dense networks. Note \citet{nandy_b_2020, saint2019using} and \citet{gui_network_2015} all assume that the network is known, where in fact it is highly possible there are unobserved covariates or network effects influencing network structure and user response. \citet{bajari2021multiple} employ the producer-consumer marketplace set-up to address interference without a network model. Rather, users are assumed to belong to a number of different populations that serve as indices for the outcomes and treatment assignments. \citet{bajari2021multiple} define a new class of experimental designs, \textit{Multiple Randomization Designs}, that model the response as a tuple with elements corresponding to each population and randomize treatments at the tuple-level. 

Despite the drawbacks of defining a parametric model for $Y_i$, there are inherit advantages to this approach, such as analyzing heterogeneity in the form of interactions or applying conventional tools like censoring and stratification \citep{walker_design_2014}. Under this framework, a natural solution to the question of treatment allocation is optimal design of experiments theory. Optimal design refers to the general practice of choosing a design matrix from the space of potential candidates, $X \in \mathcal{X}$, according to various optimality criterion. In \citet{parker_optimal_2017}, the response is modelled as $Y_i = \alpha + \tau_{t(i)} + \sum_{j=1}^n A_{ij}\gamma_{t(j)} + \eps_i$, where $\tau_{t(i)}$ represents the treatment applied to $i$, assuming $k \in \{1,...K\}$ treatments. A blocking parameter $b_i$ can also be introduced to this model \citep{koutra_designing_2017}. With this framework, \citet{parker_optimal_2017} and \citet{koutra_designing_2017} provide some interesting insights into what optimal designs for network A/B testing might look like, namely that unbalanced designs tend to be better at reducing the variance of $\hat{\tau}_j$ than balanced allocation. However, these models are rather unrealistic. Because they do not scale the spillover effect by the degree of node $i$, as the number of neighbors of $i$ grows, $\sum_{j=1}^n A_{ij}\gamma_{t(j)} \rightarrow \infty$ as well, meaning the spillover effect completely dominates $\tau_{t(i)}$ for the large networks typically observed in OCEs. \citet{parker_optimal_2017} and \citet{koutra_designing_2017} also do not optimize for the ATE, instead considering optimal designs for only $\tau_j$ by minimizing the average variance of all pairwise treatment effects. Additionally, these optimal designs are chosen with an exhaustive search algorithm, which searches the entire space of $\mathcal{X}$, or $K^n$ potential designs, before selecting $X$. Indeed, some of the designs obtained via search algorithm in \citet{parker_optimal_2017} were outperformed by randomly generating $X$. \citet{pokhiko_d-optimal_2019} and \citet{zhang_optimal_2020} alternatively choose conditional auto-regressive models to mimic the network effect by correlating the response error of $i$ with that of its neighbors. A strong limitation of this approach is this correlation is assumed to be the same across all nodes. \citet{zhang_optimal_2020} address this issue by using Bayesian priors via simulation, but do not leverage network information in defining them. 

\section{Beyond This Review}\label{sec:beyondSM}

We have presented literature that generally assumes a single treatment and control under a frequentist framework. While this setting describes an appreciable majority of OCEs, there is also growing interest in methodologies that extend beyond the scope of this review. Researchers aiming to circumvent limitations of the frequentist p-value have turned to Bayesian methods \citep{stucchio_bayesian_2015, letham2019, deng2016continuous, deng_post-selection_2019, kamalbasha2021bayesian, hoffmann2021bayesian}, including implementations of Bayes factor hypothesis testing \citep{deng_objective_2015} and tests for practical significance \citep{stevens2022comparative}. Many practitioners have noted that the ATE itself is not a quantity of interest in several applications, e.g., when optimizing tail performance, and have begun to develop approaches using \textit{quantile metrics} \citep{liu2019large, howard2019sequential, uber_quantiles_2018}.  \textit{Multi-armed bandits} have been used to handle multiple treatments in online settings, with a focus on sequential decision-making and exposing more users to successful variants to increase reward \citep{liu_trading_2014, issa_mattos_multi-armed_2019, alex_birkett_when_2019, stitch_fix_mabs, lomas_interface_2016}. Thompson sampling \citep{scott_modern_2010, scott_multi-armed_2015, dimakopoulou2021online} as well as contextual bandits \citep{li_contextual-bandit_2010, agarwal2016making} have all been used in industry. Novel experimental designs have also been developed for purposes of increasing sensitivity in low-power settings; several of these were discussed in Section 6 in the context of mitigating interference. Another commonly used design, particularly in the context of experiments on search ranking algorithms, is \textit{interleaving} \citep{microsoft2013interleaving, netflix2017interleave, airbnb2022interleave}. Rather than displaying results to a user from either a treatment algorithm or a control algorithm, this design involves interleaving search results from both the treatment and control algorithms. Thus, each user experiences both the treatment and control simultaneously, thereby providing additional information, yielding insights faster.

Although OCEs with multiple variants are reasonably common, full- and fractional-factorial experiments that emphasize estimation of main and interaction effects are uncommon; \citet{kohavi_controlled_2009, georgiev2019statmethods} argue that the added practical complexity of such experiments hurts development agility and is not worth the additional effort when interactions in practice are rare. They suggest that it is preferable to run multiple single-factor experiments concurrently, , and validate that there are no significant interactions between all pairs of experiments \citep{gupta_summit_2019}. \textit{Multivariate tests} (where the multiple variants are defined by the factorial enumeration of multiple factors' levels) \textit{do} exist in this space \citep{mcfarland_experiment_2012, wildman_using_2019}, but the goal of the analysis is primarily to identify the optimal variant, \textit{not} to estimate individual effects. Though multivariate tests are not as common as A/B or A/B/n tests, research in this area carries on \citep{sadeghi_sliced_2019}, with recent research in optimal design \citep{bhat_near-optimal_2020, basse2023minimax, bojinov2022design}, nonparametric estimators for panel experiments \citep{bojinov2021panel}, and factorial designs for sequential testing \citep{haizler_factorial_2020}. How to avoid, identify, and estimate interactions between multiple concurrent experiments is also of great interest \citep{kohavi_controlled_2009, gupta_summit_2019, overlapping_2021}.

Another important facet of OCEs outside the scope of this review is the issue of ethics \citep{gupta_summit_2019, kohavi_trustworthy_2020}. As noted, the experimental units in OCEs are often people -- human subjects -- and so a salient concern is whether experiments involving them are ethical. Many OCEs test harmless interface changes, but there exist A/B tests that through \textit{code} induce \textit{deception}, thus named C/D tests \citep{benbunan2017ethics, convert2021ethics}. One example is Facebook's infamous \textit{emotional contagion} experiment in which the sentiment of content shown in nearly 700,000 users' News Feeds was altered to determine whether this impacted their own emotions \citep{kramer2014experimental}. Another example is OKCupid's \textit{power of suggestion} experiment in which matched users were told their compatibility was higher than what the matching algorithm predicted in order to investigate the impact of simply telling couples they're a good match \citep{rudder2014okcupid}. 

More recently, there were ethical questions \citep{singer2022linkedin} about a retrospective analysis of experiments run by LinkedIn from 2015 to 2019 \citep{rajkumar2022causal} in order to understand the \textit{strength of weak ties} social theory \citep{granovetter1973strength}. These experiments engaged 20 million users and tested changes designed to improve the algorithm underlying the “People You May Know” (PYMK) feature. It is important to note that LinkedIn did not intentionally vary the proportion of weak and strong contacts suggested by PYMK \citep{belanger2022ethics} but that these variations were side effects of experiments optimizing for other criteria.  It is unknown if these changes have negatively or positively impacted users looking for job opportunities. Another context in which OCEs may have unintended side effects is digital labor platforms in today's ``gig'' economy. In this setting, the experimental units are typically the workers using the platform and researchers have found that continuous and concealed experimentation diminishes worker autonomy and satisfaction \citep{rahman2023experimental}. 

The primary concern in these settings is informed consent; users generally do not know when they're being experimented on, nor do they necessarily have a way to opt out of such an experiment. They implicitly consent to such experimentation when they agree to a service's terms and conditions, however, whether such consent is \textit{informed} is debatable  \citep{benbunan2017ethics}. Academics involved in human subjects research will be familiar with institutional review boards (IRBs) and ethics clearance. Such formal oversight is generally absent in the private sector. However, \citet{kohavi_trustworthy_2020} do advocate for the establishment of processes that fulfill this purpose so that an experiment's risks and benefits are carefully considered, and transparent protocols for informed consent and drop-out are instated. The authors also advocate for tools, infrastructure, and processes to ensure data security and data privacy, another issue especially relevant in this day and age. See \citet{kohavi_trustworthy_2020,bojinov2022online} for expanded discussions of identified data, anonymous data, re-identification, and differential privacy in the context of OCEs.

In contexts where a controlled experiment is unethical or infeasible, companies have turned to observational causal inference methods. For instance, Mozilla is interested in the impact of ad blocker installation on browser engagement \citep{miroglio2018effect}; Netflix wants to quantify the cumulative effect of in-device promotions and out-of-home marketing for a particular title \citep{netflix2018quasi}; Uber Eats is interested in understanding how delivery delays influence a user's future engagement with the platform, and Uber is interested in how ride bookings are impacted by surge pricing rates \citep{harinen2019using}. In these cases, and others like them, a traditional OCE is not appropriate or not possible, so companies estimate causal impacts using methods like matching, regression discontinuity, interrupted time series, instrumental variables, and difference in differences, among others. Like OCEs, this is a rapidly growing area that merits a literature review of its own.


\printbibliography

@misc{yu2020new,
      title={A New Framework for Online Testing of Heterogeneous Treatment Effect}, 
      author={Miao Yu and Wenbin Lu and Rui Song},
      year={2020},
      eprint={2002.03277},
      archivePrefix={arXiv},
      primaryClass={stat.ME}
}

@misc{ecommerce2022,
author = {Michael Keenan},
title = {Global Ecommerce Explained: Stats and Trends to Watch in 2022},
howpublished = {\url{https://www.shopify.ca/enterprise/global-ecommerce-statistics}},
month = {11},
year = {2022},
note = {(Accessed on 04/30/2023)}
}

@misc{digital2023,
author = {Simon Kemp},
title = {DIGITAL 2023: Global Overview Report},
howpublished = {\url{https://datareportal.com/reports/digital-2023-global-overview-report}},
month = {1},
year = {2023},
note = {(Accessed on 04/30/2023)}
}

@article{gupta_summit_2019,
author = {Gupta, Somit and Kohavi, Ronny and Tang, Diane and Xu, Ya and Andersen, Reid and Bakshy, Eytan and Cardin, Niall and Chandran, Sumita and Chen, Nanyu and Coey, Dominic and Curtis, Mike and Deng, Alex and Duan, Weitao and Forbes, Peter and Frasca, Brian and Guy, Tommy and Imbens, Guido W. and Saint Jacques, Guillaume and Kantawala, Pranav and Katsev, Ilya and Katzwer, Moshe and Konutgan, Mikael and Kunakova, Elena and Lee, Minyong and Lee, MJ and Liu, Joseph and McQueen, James and Najmi, Amir and Smith, Brent and Trehan, Vivek and Vermeer, Lukas and Walker, Toby and Wong, Jeffrey and Yashkov, Igor},
title = {Top Challenges from the First Practical Online Controlled Experiments Summit},
year = {2019},
issue_date = {June 2019},
publisher = {Association for Computing Machinery},
address = {New York, NY, USA},
volume = {21},
number = {1},
issn = {1931-0145},
url = {https://doi-org.prox.lib.ncsu.edu/10.1145/3331651.3331655},
doi = {10.1145/3331651.3331655},
journal = {SIGKDD Explor. Newsl.},
month = may,
pages = {20–35},
numpages = {16}
}

@article{chen2018automatic,
  title={Automatic Detection and Diagnosis of Biased Online Experiments},
  author={Chen, Nanyu and Liu, Min and Xu, Ya},
  journal={arXiv preprint arXiv:1808.00114},
  year={2018}
}

@misc{athey2020estimating,
      title={Estimating Treatment Effects using Multiple Surrogates: The Role of the Surrogate Score and the Surrogate Index}, 
      author={Susan Athey and Raj Chetty and Guido Imbens and Hyunseung Kang},
      year={2020},
      eprint={1603.09326},
      archivePrefix={arXiv},
      primaryClass={stat.ME}
}

@article{begg2000use,
  title={On the use of surrogate end points in randomized trials},
  author={Begg, Colin B and Leung, Denis HY},
  journal={Journal of the Royal Statistical Society: Series A (Statistics in Society)},
  volume={163},
  number={1},
  pages={15--28},
  year={2000},
  publisher={Wiley Online Library}
}

@article{frangakis2002principal,
  title={Principal stratification in causal inference},
  author={Frangakis, Constantine E and Rubin, Donald B},
  journal={Biometrics},
  volume={58},
  number={1},
  pages={21--29},
  year={2002},
  publisher={Wiley Online Library}
}

@article{shi2020reinforcement,
  title={A Reinforcement Learning Framework for Time-Dependent Causal Effects Evaluation in A/B Testing},
  author={Shi, Chengchun and Wang, Xiaoyu and Luo, Shikai and Song, Rui and Zhu, Hongtu and Ye, Jieping},
  journal={arXiv preprint arXiv:2002.01711},
  year={2020}
}

@inproceedings{deng2015diluted,
  title={Diluted Treatment Effect Estimation for Trigger Analysis in Online Controlled Experiments},
  author={Deng, Alex and Hu, Victor},
  booktitle={Proceedings of the Eighth ACM International Conference on Web Search and Data Mining},
  pages={349--358},
  year={2015}
}

@inproceedings{dmitriev2016pitfalls,
  title={Pitfalls of long-term online controlled experiments},
  author={Dmitriev, Pavel and Frasca, Brian and Gupta, Somit and Kohavi, Ronny and Vaz, Garnet},
  booktitle={2016 IEEE international conference on big data (big data)},
  pages={1367--1376},
  year={2016},
  organization={IEEE}
}

@article{prentice1989surrogate,
  title={Surrogate endpoints in clinical trials: definition and operational criteria},
  author={Prentice, Ross L},
  journal={Statistics in medicine},
  volume={8},
  number={4},
  pages={431--440},
  year={1989},
  publisher={Wiley Online Library}
}

@inproceedings{hassan2013beyond,
  title={Beyond clicks: query reformulation as a predictor of search satisfaction},
  author={Hassan, Ahmed and Shi, Xiaolin and Craswell, Nick and Ramsey, Bill},
  booktitle={Proceedings of the 22nd ACM international conference on Information \& Knowledge Management},
  pages={2019--2028},
  year={2013}
}

@article{cheng2020long,
  title={Long-Term Effect Estimation with Surrogate Representation},
  author={Cheng, Lu and Guo, Ruocheng and Liu, Huan},
  journal={arXiv preprint arXiv:2008.08236},
  year={2020}
}

@article{ensor2016statistical,
  title={Statistical approaches for evaluating surrogate outcomes in clinical trials: a systematic review},
  author={Ensor, Hannah and Lee, Robert J and Sudlow, Cathie and Weir, Christopher J},
  journal={Journal of biopharmaceutical statistics},
  volume={26},
  number={5},
  pages={859--879},
  year={2016},
  publisher={Taylor \& Francis}
}

@inproceedings{tran2019learning,
  title={Learning triggers for heterogeneous treatment effects},
  author={Tran, Christopher and Zheleva, Elena},
  booktitle={Proceedings of the AAAI Conference on Artificial Intelligence},
  volume={33},
  pages={5183--5190},
  year={2019}
}

@article{wagner_heteroRF_2018,
author = {Stefan Wager and Susan Athey},
title = {Estimation and Inference of Heterogeneous Treatment Effects using Random Forests},
journal = {Journal of the American Statistical Association},
volume = {113},
number = {523},
pages = {1228-1242},
year  = {2018},
publisher = {Taylor & Francis},
doi = {10.1080/01621459.2017.1319839},
URL = {https://doi.org/10.1080/01621459.2017.1319839},
eprint = {https://doi.org/10.1080/01621459.2017.1319839}
}

@article {athey_heteroPartitioning_2016,
	author = {Athey, Susan and Imbens, Guido},
	title = {Recursive partitioning for heterogeneous causal effects},
	volume = {113},
	number = {27},
	pages = {7353--7360},
	year = {2016},
	doi = {10.1073/pnas.1510489113},
	publisher = {National Academy of Sciences},
	issn = {0027-8424},
	URL = {https://www.pnas.org/content/113/27/7353},
	eprint = {https://www.pnas.org/content/113/27/7353.full.pdf},
	journal = {Proceedings of the National Academy of Sciences}
}

@article{agarwal2016making,
  title={Making contextual decisions with low technical debt},
  author={Agarwal, Alekh and Bird, Sarah and Cozowicz, Markus and Hoang, Luong and Langford, John and Lee, Stephen and Li, Jiaji and Melamed, Dan and Oshri, Gal and Ribas, Oswaldo and others},
  journal={arXiv preprint arXiv:1606.03966},
  year={2016}
}

@article{letham2019,
author = "Letham, Benjamin and Karrer, Brian and Ottoni, Guilherme and Bakshy, Eytan",
doi = "10.1214/18-BA1110",
fjournal = "Bayesian Analysis",
journal = "Bayesian Anal.",
month = "06",
number = "2",
pages = "495--519",
publisher = "International Society for Bayesian Analysis",
title = "Constrained Bayesian Optimization with Noisy Experiments",
url = "https://doi.org/10.1214/18-BA1110",
volume = "14",
year = "2019"
}

@article{yingqi_2012,
author = { Yingqi   Zhao  and  Donglin   Zeng  and  A. John   Rush  and  Michael R.   Kosorok },
title = {Estimating Individualized Treatment Rules Using Outcome Weighted Learning},
journal = {Journal of the American Statistical Association},
volume = {107},
number = {499},
pages = {1106-1118},
year  = {2012},
publisher = {Taylor & Francis},
doi = {10.1080/01621459.2012.695674},
note ={PMID: 23630406},
URL = {https://doi.org/10.1080/01621459.2012.695674},
eprint = {https://doi.org/10.1080/01621459.2012.695674}
}

@inproceedings{wang2019heavy,
  title={On Heavy-user Bias in A/B Testing},
  author={Wang, Yu and Gupta, Somit and Lu, Jiannan and Mahmoudzadeh, Ali and Liu, Sophia},
  booktitle={Proceedings of the 28th ACM International Conference on Information and Knowledge Management},
  pages={2425--2428},
  year={2019}
}

@article{barber2015controlling,
  title={Controlling the false discovery rate via knockoffs},
  author={Barber, Rina Foygel and Cand{\`e}s, Emmanuel J and others},
  journal={The Annals of Statistics},
  volume={43},
  number={5},
  pages={2055--2085},
  year={2015},
  publisher={Institute of Mathematical Statistics}
}

@inproceedings{syrgkanis2019machine,
  title={Machine learning estimation of heterogeneous treatment effects with instruments},
  author={Syrgkanis, Vasilis and Lei, Victor and Oprescu, Miruna and Hei, Maggie and Battocchi, Keith and Lewis, Greg},
  booktitle={Advances in Neural Information Processing Systems},
  pages={15193--15202},
  year={2019}
}

@article{deng2016concise,
  title={Concise summarization of heterogeneous treatment effect using total variation regularized regression},
  author={Deng, Alex and Zhang, Pengchuan and Chen, Shouyuan and Kim, Dong Woo and Lu, Jiannan},
  journal={arXiv preprint arXiv:1610.03917},
  year={2016}
}

@inproceedings{deng2016continuous,
  title={Continuous monitoring of A/B tests without pain: Optional stopping in Bayesian testing},
  author={Deng, Alex and Lu, Jiannan and Chen, Shouyuan},
  booktitle={2016 IEEE International Conference on Data Science and Advanced Analytics (DSAA)},
  pages={243--252},
  year={2016},
  organization={IEEE}
}

@inproceedings{kharitonov2017_learning,
author = {Kharitonov, Eugene and Drutsa, Alexey and Serdyukov, Pavel},
title = {Learning Sensitive Combinations of A/B Test Metrics},
year = {2017},
isbn = {9781450346757},
publisher = {Association for Computing Machinery},
address = {New York, NY, USA},
url = {https://doi-org.prox.lib.ncsu.edu/10.1145/3018661.3018708},
doi = {10.1145/3018661.3018708},
booktitle = {Proceedings of the Tenth ACM International Conference on Web Search and Data Mining},
pages = {651–659},
numpages = {9},
location = {Cambridge, United Kingdom},
series = {WSDM '17}
}

@inproceedings{poyarkov2016boosted,
author = {Poyarkov, Alexey and Drutsa, Alexey and Khalyavin, Andrey and Gusev, Gleb and Serdyukov, Pavel},
title = {Boosted Decision Tree Regression Adjustment for Variance Reduction in Online Controlled Experiments},
year = {2016},
isbn = {9781450342322},
publisher = {Association for Computing Machinery},
address = {New York, NY, USA},
url = {https://doi-org.prox.lib.ncsu.edu/10.1145/2939672.2939688},
doi = {10.1145/2939672.2939688},
booktitle = {Proceedings of the 22nd ACM SIGKDD International Conference on Knowledge Discovery and Data Mining},
pages = {235–244},
numpages = {10},
location = {San Francisco, California, USA},
series = {KDD '16}
}

@inproceedings{drutsa2015future,
  title={Future user engagement prediction and its application to improve the sensitivity of online experiments},
  author={Drutsa, Alexey and Gusev, Gleb and Serdyukov, Pavel},
  booktitle={Proceedings of the 24th International Conference on World Wide Web},
  pages={256--266},
  year={2015}
}

@inproceedings{drusta2015practical,
author = {Drutsa, Alexey and Ufliand, Anna and Gusev, Gleb},
title = {Practical Aspects of Sensitivity in Online Experimentation with User Engagement Metrics},
year = {2015},
isbn = {9781450337946},
publisher = {Association for Computing Machinery},
address = {New York, NY, USA},
url = {https://doi-org.prox.lib.ncsu.edu/10.1145/2806416.2806496},
doi = {10.1145/2806416.2806496},
booktitle = {Proceedings of the 24th ACM International on Conference on Information and Knowledge Management},
pages = {763–772},
numpages = {10},
location = {Melbourne, Australia},
series = {CIKM '15}
}

@inproceedings{kharitonov2015sequential,
author = {Kharitonov, Eugene and Vorobev, Aleksandr and Macdonald, Craig and Serdyukov, Pavel and Ounis, Iadh},
title = {Sequential Testing for Early Stopping of Online Experiments},
year = {2015},
isbn = {9781450336215},
publisher = {Association for Computing Machinery},
address = {New York, NY, USA},
url = {https://doi-org.prox.lib.ncsu.edu/10.1145/2766462.2767729},
doi = {10.1145/2766462.2767729},
booktitle = {Proceedings of the 38th International ACM SIGIR Conference on Research and Development in Information Retrieval},
pages = {473–482},
numpages = {10},
location = {Santiago, Chile},
series = {SIGIR '15}
}

@article{acharya2013sampling,
  title={Sampling: Why and how of it},
  author={Acharya, Anita S and Prakash, Anupam and Saxena, Pikee and Nigam, Aruna},
  journal={Indian Journal of Medical Specialties},
  volume={4},
  number={2},
  pages={330--333},
  year={2013}
}

@misc{HowBook_online,
author = {Simon Jackson},
title = {How Booking.com increases the power of online experiments with CUPED | Booking.com Data Science},
howpublished = {\url{https://booking.ai/how-booking-com-increases-the-power-of-online-experiments-with-cuped-995d186fff1d}},
month = {1},
year = {2018},
note = {(Accessed on 01/13/2021)}
}

@inproceedings{liou2020variance,
  title={Variance-Weighted Estimators to Improve Sensitivity in Online Experiments},
  author={Liou, Kevin and Taylor, Sean J},
  booktitle={Proceedings of the 21st ACM Conference on Economics and Computation},
  pages={837--850},
  year={2020}
}

@article{saint2019using,
  title={Using Ego-Clusters to Measure Network Effects at LinkedIn},
  author={Saint-Jacques, Guillaume and Varshney, Maneesh and Simpson, Jeremy and Xu, Ya},
  journal={arXiv preprint arXiv:1903.08755},
  year={2019}
}

@article{egami2017unbiased,
  title={Unbiased estimation and sensitivity analysis for network-specific spillover effects: Application to an online network experiment},
  author={Egami, Naoki},
  journal={arXiv preprint arXiv:1708.08171},
  year={2017}
}

@article{wald_1945_sequential,
 ISSN = {00034851},
 URL = {http://www.jstor.org/stable/2235829},
 author = {A. Wald},
 journal = {The Annals of Mathematical Statistics},
 number = {2},
 pages = {117--186},
 publisher = {Institute of Mathematical Statistics},
 title = {Sequential Tests of Statistical Hypotheses},
 volume = {16},
 year = {1945}
}

@inproceedings{ju_2019_sequential_ab_test,
author = {Ju, Nianqiao and Hu, Diane and Henderson, Adam and Hong, Liangjie},
title = {A Sequential Test for Selecting the Better Variant: Online A/B Testing, Adaptive Allocation, and Continuous Monitoring},
year = {2019},
isbn = {9781450359405},
publisher = {Association for Computing Machinery},
address = {New York, NY, USA},
url = {https://doi-org.prox.lib.ncsu.edu/10.1145/3289600.3291025},
doi = {10.1145/3289600.3291025},
booktitle = {Proceedings of the Twelfth ACM International Conference on Web Search and Data Mining},
pages = {492–500},
numpages = {9},
location = {Melbourne VIC, Australia},
series = {WSDM '19}
}

@inproceedings{abhishek_2017_nonparametric_sequential,
author = {Abhishek, Vineet and Mannor, Shie},
title = {A Nonparametric Sequential Test for Online Randomized Experiments},
year = {2017},
isbn = {9781450349147},
publisher = {International World Wide Web Conferences Steering Committee},
address = {Republic and Canton of Geneva, CHE},
url = {https://doi-org.prox.lib.ncsu.edu/10.1145/3041021.3054196},
doi = {10.1145/3041021.3054196},
booktitle = {Proceedings of the 26th International Conference on World Wide Web Companion},
pages = {610–616},
numpages = {7},
location = {Perth, Australia},
series = {WWW '17 Companion}
}

@article{chan_gsprt_2005,
author = {H.P. Chan and T.L. Lai},
title = {Importance Sampling for Generalized Likelihood Ratio Procedures in Sequential Analysis},
journal = {Sequential Analysis},
volume = {24},
number = {3},
pages = {259-278},
year  = {2005},
publisher = {Taylor & Francis},
doi = {10.1081/SQA-200063280},

URL = { 
        https://doi.org/10.1081/SQA-200063280
    
},
eprint = { 
        https://doi.org/10.1081/SQA-200063280
    
}

}

@article{robinson_1988,
 ISSN = {00129682, 14680262},
 URL = {http://www.jstor.org/stable/1912705},
 author = {P. M. Robinson},
 journal = {Econometrica},
 number = {4},
 pages = {931--954},
 publisher = {[Wiley, Econometric Society]},
 title = {Root-N-Consistent Semiparametric Regression},
 volume = {56},
 year = {1988}
}

@article{imai_hte_2013,
author = {Kosuke Imai and Marc Ratkovic},
title = {{Estimating treatment effect heterogeneity in randomized program evaluation}},
volume = {7},
journal = {The Annals of Applied Statistics},
number = {1},
publisher = {Institute of Mathematical Statistics},
pages = {443 -- 470},
year = {2013},
doi = {10.1214/12-AOAS593},
URL = {https://doi.org/10.1214/12-AOAS593}
}

@article{chernozhukov2017double,
  title={Double/debiased/neyman machine learning of treatment effects},
  author={Chernozhukov, Victor and Chetverikov, Denis and Demirer, Mert and Duflo, Esther and Hansen, Christian and Newey, Whitney},
  journal={American Economic Review},
  volume={107},
  number={5},
  pages={261--65},
  year={2017}
}

@article{kohavi_controlled_2009,
	title = {Controlled experiments on the web: survey and practical guide},
	volume = {18},
	issn = {1384-5810, 1573-756X},
	url = {http://link.springer.com/10.1007/s10618-008-0114-1},
	doi = {10.1007/s10618-008-0114-1},
	shorttitle = {Controlled experiments on the web},
	pages = {140--181},
	number = {1},
	journaltitle = {Data Mining and Knowledge Discovery},
	shortjournal = {Data Min Knowl Disc},
	author = {Kohavi, Ronny and Longbotham, Roger and Sommerfield, Dan and Henne, Randal M.},
	urldate = {2020-05-12},
	date = {2009-02},
	langid = {english}
}

@article{haizler_factorial_2020,
	title = {Factorial Designs for Online Experiments},
	issn = {0040-1706, 1537-2723},
	url = {https://www.tandfonline.com/doi/full/10.1080/00401706.2019.1701556},
	doi = {10.1080/00401706.2019.1701556},
	pages = {1--12},
	journaltitle = {Technometrics},
	shortjournal = {Technometrics},
	author = {Haizler, Tamar and Steinberg, David M.},
	urldate = {2020-05-12},
	date = {2020-01-23},
	langid = {english}
}

@article{kohavi_surprising_2017,
	title = {The Surprising Power of Online Experiments},
	issn = {0017-8012},
	url = {https://hbr.org/2017/09/the-surprising-power-of-online-experiments},
	issue = {September–October 2017},
	journaltitle = {Harvard Business Review},
	author = {Kohavi, Ronny and Thomke, Stefan},
	urldate = {2020-05-12},
	date = {2017-09-01}
}

@article{pokhiko_d-optimal_2019,
	title = {D-optimal Design for Network A/B Testing},
	volume = {13},
	issn = {1559-8608, 1559-8616},
	url = {http://arxiv.org/abs/1902.00482},
	doi = {10.1007/s42519-019-0058-3},
	pages = {61},
	number = {4},
	journaltitle = {Journal of Statistical Theory and Practice},
	shortjournal = {J Stat Theory Pract},
	author = {Pokhiko, Victoria and Zhang, Qiong and Kang, Lulu and Mays, D'arcy P.},
	urldate = {2020-05-13},
	date = {2019-12},
	eprinttype = {arxiv},
	eprint = {1902.00482}
}

@article{sadeghi_sliced_2019,
	title = {Sliced Designs for Multi-Platform Online Experiments},
	issn = {0040-1706},
	url = {https://amstat.tandfonline.com/doi/abs/10.1080/00401706.2019.1647288},
	doi = {10.1080/00401706.2019.1647288},
	pages = {1--16},
	journaltitle = {Technometrics},
	shortjournal = {Technometrics},
	author = {Sadeghi, Soheil and Chien, Peter and Arora, Neeraj},
	urldate = {2020-05-13},
	date = {2019-08-14},
	note = {Publisher: Taylor \& Francis}
}

@inproceedings{gui_network_2015,
	location = {Florence, Italy},
	title = {Network A/B Testing: From Sampling to Estimation},
	isbn = {978-1-4503-3469-3},
	url = {https://doi.org/10.1145/2736277.2741081},
	doi = {10.1145/2736277.2741081},
	series = {{WWW} '15},
	shorttitle = {Network A/B Testing},
	pages = {399--409},
	booktitle = {Proceedings of the 24th International Conference on World Wide Web},
	publisher = {International World Wide Web Conferences Steering Committee},
	author = {Gui, Huan and Xu, Ya and Bhasin, Anmol and Han, Jiawei},
	urldate = {2020-05-14},
	date = {2015-05-18}
}

@inproceedings{xu_infrastructure_2015,
	location = {Sydney, {NSW}, Australia},
	title = {From Infrastructure to Culture: A/B Testing Challenges in Large Scale Social Networks},
	isbn = {978-1-4503-3664-2},
	url = {http://dl.acm.org/citation.cfm?doid=2783258.2788602},
	doi = {10.1145/2783258.2788602},
	shorttitle = {From Infrastructure to Culture},
	eventtitle = {the 21th {ACM} {SIGKDD} International Conference},
	pages = {2227--2236},
	booktitle = {Proceedings of the 21th {ACM} {SIGKDD} International Conference on Knowledge Discovery and Data Mining - {KDD} '15},
	publisher = {{ACM} Press},
	author = {Xu, Ya and Chen, Nanyu and Fernandez, Addrian and Sinno, Omar and Bhasin, Anmol},
	urldate = {2020-05-14},
	date = {2015},
	langid = {english}
}

@inproceedings{kohavi_online_2013,
	location = {Chicago, Illinois, {USA}},
	title = {Online controlled experiments at large scale},
	isbn = {978-1-4503-2174-7},
	url = {http://dl.acm.org/citation.cfm?doid=2487575.2488217},
	doi = {10.1145/2487575.2488217},
	eventtitle = {the 19th {ACM} {SIGKDD} international conference},
	pages = {1168},
	booktitle = {Proceedings of the 19th {ACM} {SIGKDD} international conference on Knowledge discovery and data mining - {KDD} '13},
	publisher = {{ACM} Press},
	author = {Kohavi, Ronny and Deng, Alex and Frasca, Brian and Walker, Toby and Xu, Ya and Pohlmann, Nils},
	urldate = {2020-05-14},
	date = {2013},
	langid = {english}
}

@inproceedings{kohavi_seven_2014,
	location = {New York, New York, {USA}},
	title = {Seven rules of thumb for web site experimenters},
	isbn = {978-1-4503-2956-9},
	url = {http://dl.acm.org/citation.cfm?doid=2623330.2623341},
	doi = {10.1145/2623330.2623341},
	eventtitle = {the 20th {ACM} {SIGKDD} international conference},
	pages = {1857--1866},
	booktitle = {Proceedings of the 20th {ACM} {SIGKDD} international conference on Knowledge discovery and data mining - {KDD} '14},
	publisher = {{ACM} Press},
	author = {Kohavi, Ronny and Deng, Alex and Longbotham, Roger and Xu, Ya},
	urldate = {2020-05-14},
	date = {2014},
	langid = {english}
}

@inproceedings{crook_seven_2009,
	location = {Paris, France},
	title = {Seven pitfalls to avoid when running controlled experiments on the web},
	isbn = {978-1-60558-495-9},
	url = {http://portal.acm.org/citation.cfm?doid=1557019.1557139},
	doi = {10.1145/1557019.1557139},
	eventtitle = {the 15th {ACM} {SIGKDD} international conference},
	pages = {1105},
	booktitle = {Proceedings of the 15th {ACM} {SIGKDD} international conference on Knowledge discovery and data mining - {KDD} '09},
	publisher = {{ACM} Press},
	author = {Crook, Thomas and Frasca, Brian and Kohavi, Ronny and Longbotham, Roger},
	urldate = {2020-05-14},
	date = {2009},
	langid = {english}
}

@inproceedings{xie_false_2018,
	location = {London, United Kingdom},
	title = {False Discovery Rate Controlled Heterogeneous Treatment Effect Detection for Online Controlled Experiments},
	isbn = {978-1-4503-5552-0},
	url = {https://doi.org/10.1145/3219819.3219860},
	doi = {10.1145/3219819.3219860},
	series = {{KDD} '18},
	pages = {876--885},
	booktitle = {Proceedings of the 24th {ACM} {SIGKDD} International Conference on Knowledge Discovery \& Data Mining},
	publisher = {Association for Computing Machinery},
	author = {Xie, Yuxiang and Chen, Nanyu and Shi, Xiaolin},
	urldate = {2020-05-20},
	date = {2018-07-19}
}

@inproceedings{tang_overlapping_2010,
	location = {Washington, {DC}, {USA}},
	title = {Overlapping experiment infrastructure: more, better, faster experimentation},
	isbn = {978-1-4503-0055-1},
	url = {https://doi.org/10.1145/1835804.1835810},
	doi = {10.1145/1835804.1835810},
	series = {{KDD} '10},
	shorttitle = {Overlapping experiment infrastructure},
	pages = {17--26},
	booktitle = {Proceedings of the 16th {ACM} {SIGKDD} international conference on Knowledge discovery and data mining},
	publisher = {Association for Computing Machinery},
	author = {Tang, Diane and Agarwal, Ashish and O'Brien, Deirdre and Meyer, Mike},
	urldate = {2020-05-20},
	date = {2010-07-25}
}

@book{kohavi_trustworthy_2020,
	location = {Cambridge},
	title = {Trustworthy Online Controlled Experiments: A Practical Guide to A/B Testing},
	isbn = {978-1-108-72426-5},
	shorttitle = {Trustworthy Online Controlled Experiments},
	publisher = {Cambridge University Press},
	author = {Kohavi, Ronny and Tang, Diane and Xu, Ya},
	date = {2020},
        doi = {10.1017/9781108653985},
        note = {\url{https://experimentguide.com/}}
}

@book{mcfarland_experiment_2012,
	title = {Experiment!: Website conversion rate optimization with A/B and multivariate testing},
	isbn = {978-0-13-304008-1},
	shorttitle = {Experiment!},
	pagetotal = {190},
	publisher = {New Riders},
	author = {{McFarland}, Colin},
	date = {2012-08-17},
	langid = {english}
}

@inproceedings{van2023estimating,
  title={Estimating long-term causal effects from short-term experiments and long-term observational data with unobserved confounding},
  author={Van Goffrier, Graham and Maystre, Lucas and Gilligan-Lee, Ciar{\'a}n},
  journal={arXiv preprint arXiv:2302.10625},
  year={2023},
	pages = {786--794},
	booktitle = {Proceedings of Machine Learning Research},
	volume = {213}
}

@inproceedings{kohavi_trustworthy_2012,
	location = {Beijing, China},
	title = {Trustworthy online controlled experiments: five puzzling outcomes explained},
	isbn = {978-1-4503-1462-6},
	url = {https://doi.org/10.1145/2339530.2339653},
	doi = {10.1145/2339530.2339653},
	series = {{KDD} '12},
	shorttitle = {Trustworthy online controlled experiments},
	pages = {786--794},
	booktitle = {Proceedings of the 18th {ACM} {SIGKDD} international conference on Knowledge discovery and data mining},
	publisher = {Association for Computing Machinery},
	author = {Kohavi, Ronny and Deng, Alex and Frasca, Brian and Longbotham, Roger and Walker, Toby and Xu, Ya},
	urldate = {2020-05-21},
	date = {2012-08-12}
}

@article{bhat_near-optimal_2020,
	title = {Near-Optimal A-B Testing},
	issn = {0025-1909},
	url = {https://pubsonline.informs.org/doi/abs/10.1287/mnsc.2019.3424},
	doi = {10.1287/mnsc.2019.3424},
	journaltitle = {Management Science},
	shortjournal = {Management Science},
	author = {Bhat, Nikhil and Farias, Vivek F. and Moallemi, Ciamac C. and Sinha, Deeksha},
	urldate = {2020-05-21},
	date = {2020-04-09},
	note = {Publisher: {INFORMS}}
}

@article{johari2022always,
  title={Always valid inference: Continuous monitoring of a/b tests},
  author={Johari, Ramesh and Koomen, Pete and Pekelis, Leonid and Walsh, David},
  journal={Operations Research},
  volume={70},
  number={3},
  pages={1806--1821},
  year={2022},
  publisher={INFORMS}
}

@inproceedings{xie_improving_2016,
	location = {San Francisco, California, {USA}},
	title = {Improving the Sensitivity of Online Controlled Experiments: Case Studies at Netflix},
	isbn = {978-1-4503-4232-2},
	url = {https://doi.org/10.1145/2939672.2939733},
	doi = {10.1145/2939672.2939733},
	series = {{KDD} '16},
	shorttitle = {Improving the Sensitivity of Online Controlled Experiments},
	pages = {645--654},
	booktitle = {Proceedings of the 22nd {ACM} {SIGKDD} International Conference on Knowledge Discovery and Data Mining},
	publisher = {Association for Computing Machinery},
	author = {Xie, Huizhi and Aurisset, Juliette},
	urldate = {2020-05-21},
	date = {2016-08-13}
}

@inproceedings{deng_improving_2013,
	location = {Rome, Italy},
	title = {Improving the sensitivity of online controlled experiments by utilizing pre-experiment data},
	isbn = {978-1-4503-1869-3},
	url = {http://dl.acm.org/citation.cfm?doid=2433396.2433413},
	doi = {10.1145/2433396.2433413},
	eventtitle = {the sixth {ACM} international conference},
	pages = {123},
	booktitle = {Proceedings of the sixth {ACM} international conference on Web search and data mining - {WSDM} '13},
	publisher = {{ACM} Press},
	author = {Deng, Alex and Xu, Ya and Kohavi, Ronny and Walker, Toby},
	urldate = {2020-05-27},
	date = {2013},
	langid = {english}
}

@inproceedings{liu_trading_2014,
	title = {Trading Off Scientific Knowledge and User Learning with Multi-Armed Bandits},
	booktitle = {{EDM}},
	author = {Liu, Yun-En and Mandel, Travis and Brunskill, Emma and Popovic, Zoran},
	date = {2014}
}

@inproceedings{deng_data-driven_2016,
	location = {San Francisco California {USA}},
	title = {Data-Driven Metric Development for Online Controlled Experiments: Seven Lessons Learned},
	isbn = {978-1-4503-4232-2},
	url = {https://dl.acm.org/doi/10.1145/2939672.2939700},
	doi = {10.1145/2939672.2939700},
	shorttitle = {Data-Driven Metric Development for Online Controlled Experiments},
	eventtitle = {{KDD} '16: The 22nd {ACM} {SIGKDD} International Conference on Knowledge Discovery and Data Mining},
	pages = {77--86},
	booktitle = {Proceedings of the 22nd {ACM} {SIGKDD} International Conference on Knowledge Discovery and Data Mining},
	publisher = {{ACM}},
	author = {Deng, Alex and Shi, Xiaolin},
	urldate = {2020-05-29},
	date = {2016-08-13},
	langid = {english}
}

@inproceedings{deng_trustworthy_2017,
	location = {Cambridge, United Kingdom},
	title = {Trustworthy Analysis of Online A/B Tests: Pitfalls, challenges and solutions},
	isbn = {978-1-4503-4675-7},
	url = {https://doi.org/10.1145/3018661.3018677},
	doi = {10.1145/3018661.3018677},
	series = {{WSDM} '17},
	shorttitle = {Trustworthy Analysis of Online A/B Tests},
	pages = {641--649},
	booktitle = {Proceedings of the Tenth {ACM} International Conference on Web Search and Data Mining},
	publisher = {Association for Computing Machinery},
	author = {Deng, Alex and Lu, Jiannan and Litz, Jonthan},
	urldate = {2020-06-01},
	date = {2017-02-02}
}

@inproceedings{deng_objective_2015,
	location = {Florence, Italy},
	title = {Objective Bayesian Two Sample Hypothesis Testing for Online Controlled Experiments},
	isbn = {978-1-4503-3473-0},
	url = {https://doi.org/10.1145/2740908.2742563},
	doi = {10.1145/2740908.2742563},
	series = {{WWW} '15 Companion},
	pages = {923--928},
	booktitle = {Proceedings of the 24th International Conference on World Wide Web},
	publisher = {Association for Computing Machinery},
	author = {Deng, Alex},
	urldate = {2020-06-02},
	date = {2015-05-18}
}

@article{stucchio_bayesian_2015,
	title = {Bayesian A/B Testing at {VWO}},
	pages = {33},
	author = {Stucchio, Chris},
	date = {2015-09-02},
	langid = {english}
}

@article{christian_b_2012,
	title = {The A/B Test: Inside the Technology That's Changing the Rules of Business},
	volume = {20},
	issn = {1059-1028},
	url = {https://www.wired.com/2012/04/ff-abtesting/},
	shorttitle = {The A/B Test},
	number = {5},
	journaltitle = {Wired},
	author = {Christian, Brian},
	urldate = {2020-06-04},
	date = {2012-04-26}
}

@inproceedings{deng_post-selection_2019,
title={On Post-selection Inference in A/B Testing},
author={Deng, Alex and Li, Yicheng and Lu, Jiannan and Ramamurthy, Vivek},
booktitle={Proceedings of the 27th ACM SIGKDD Conference on Knowledge Discovery \& Data Mining},
pages={2743--2752},
year={2021}
}

@article{eckles_design_2014,
	title = {Design and analysis of experiments in networks: Reducing bias from interference},
	url = {http://arxiv.org/abs/1404.7530},
	shorttitle = {Design and analysis of experiments in networks},
	journaltitle = {{arXiv}:1404.7530 [physics, stat]},
	author = {Eckles, Dean and Karrer, Brian and Ugander, Johan},
	urldate = {2020-06-10},
	date = {2014-08-13},
	eprinttype = {arxiv},
	eprint = {1404.7530}
}

@article{walker_design_2014,
	title = {Design of Randomized Experiments in Networks},
	volume = {102},
	issn = {1558-2256},
	doi = {10.1109/JPROC.2014.2363674},
	pages = {1940--1951},
	number = {12},
	journaltitle = {Proceedings of the {IEEE}},
	author = {Walker, Dylan and Muchnik, Lev},
	date = {2014-12},
	note = {Conference Name: Proceedings of the {IEEE}}
}

@article{ugander_graph_2013,
	title = {Graph cluster randomization: network exposure to multiple universes},
	url = {http://arxiv.org/abs/1305.6979},
	shorttitle = {Graph cluster randomization},
	journaltitle = {{arXiv}:1305.6979 [physics, stat]},
	author = {Ugander, Johan and Karrer, Brian and Backstrom, Lars and Kleinberg, Jon},
	urldate = {2020-06-10},
	date = {2013-05-29},
	eprinttype = {arxiv},
	eprint = {1305.6979}
}

@article{nandy_b_2020,
  title={A/B testing in dense large-scale networks: design and inference},
  author={Nandy, Preetam and Basu, Kinjal and Chatterjee, Shaunak and Tu, Ye},
  journal={Advances in Neural Information Processing Systems},
  volume={33},
  pages={2870--2880},
  year={2020}
}

@online{sangho_yoon_designing_2018,
	title = {Designing A/B tests in a collaboration network},
	url = {http://www.unofficialgoogledatascience.com/2018/01/designing-ab-tests-in-collaboration.html},
	titleaddon = {The Unofficial Google Data Science Blog},
	author = {{Sangho Yoon}},
	urldate = {2020-06-11},
	date = {2018-01-16},
	note = {Library Catalog: www.unofficialgoogledatascience.com}
}

@inproceedings{saveski_detecting_2017,
	location = {Halifax, {NS}, Canada},
	title = {Detecting Network Effects: Randomizing Over Randomized Experiments},
	isbn = {978-1-4503-4887-4},
	url = {https://doi.org/10.1145/3097983.3098192},
	doi = {10.1145/3097983.3098192},
	series = {{KDD} '17},
	shorttitle = {Detecting Network Effects},
	pages = {1027--1035},
	booktitle = {Proceedings of the 23rd {ACM} {SIGKDD} International Conference on Knowledge Discovery and Data Mining},
	publisher = {Association for Computing Machinery},
	author = {Saveski, Martin and Pouget-Abadie, Jean and Saint-Jacques, Guillaume and Duan, Weitao and Ghosh, Souvik and Xu, Ya and Airoldi, Edoardo M.},
	urldate = {2020-06-15},
	date = {2017-08-13}
}

@article{basse_limitations_2018,
	title = {Limitations of Design-based Causal Inference and A/B Testing under Arbitrary and Network Interference},
	volume = {48},
	issn = {0081-1750},
	url = {https://doi.org/10.1177/0081175018782569},
	doi = {10.1177/0081175018782569},
	pages = {136--151},
	number = {1},
	journaltitle = {Sociological Methodology},
	shortjournal = {Sociological Methodology},
	author = {Basse, Guillaume W. and Airoldi, Edoardo M.},
	urldate = {2020-06-15},
	date = {2018-08-01},
	langid = {english},
	note = {Publisher: {SAGE} Publications Inc}
}

@article{scott_multi-armed_2015,
	title = {Multi-armed bandit experiments in the online service economy},
	volume = {31},
	issn = {1524-1904},
	url = {https://doi.org/10.1002/asmb.2104},
	doi = {10.1002/asmb.2104},
	pages = {37--45},
	number = {1},
	journaltitle = {Applied Stochastic Models in Business and Industry},
	shortjournal = {Appl. Stoch. Model. Bus. Ind.},
	author = {Scott, Steven L.},
	urldate = {2020-06-16},
	date = {2015-01-01}
}

@article{scott_modern_2010,
	title = {A modern Bayesian look at the multi-armed bandit},
	volume = {26},
	issn = {1524-1904},
	url = {https://doi.org/10.1002/asmb.874},
	doi = {10.1002/asmb.874},
	pages = {639--658},
	number = {6},
	journaltitle = {Applied Stochastic Models in Business and Industry},
	shortjournal = {Appl. Stoch. Model. Bus. Ind.},
	author = {Scott, Steven L.},
	urldate = {2020-06-16},
	date = {2010-11-01}
}

@online{alex_birkett_when_2019,
	title = {When to Run Bandit Tests Instead of A/B/n Tests},
	url = {https://cxl.com/blog/bandit-tests/},
	titleaddon = {{CXL}},
	author = {Birkett, Alex},
	urldate = {2020-06-16},
	date = {2019-10-07},
	langid = {american},
	note = {Library Catalog: cxl.com}
}

@inproceedings{hohnhold_focusing_2015,
	location = {Sydney, {NSW}, Australia},
	title = {Focusing on the Long-term: It's Good for Users and Business},
	isbn = {978-1-4503-3664-2},
	url = {https://doi.org/10.1145/2783258.2788583},
	doi = {10.1145/2783258.2788583},
	series = {{KDD} '15},
	shorttitle = {Focusing on the Long-term},
	pages = {1849--1858},
	booktitle = {Proceedings of the 21th {ACM} {SIGKDD} International Conference on Knowledge Discovery and Data Mining},
	publisher = {Association for Computing Machinery},
	author = {Hohnhold, Henning and O'Brien, Deirdre and Tang, Diane},
	urldate = {2020-06-17},
	date = {2015-08-10}
}

@online{wildman_using_2019,
	title = {Using A/B Testing, Factorial Design, and Multivariate Tests for Deep Visitor Insights},
	url = {https://www.thecreativemomentum.com/blog/using-a/b-testing-factorial-design-and-multivariate-tests-for-deep-visitor-insights},
	author = {Wildman, Brent},
	urldate = {2020-07-13},
	date = {2019-01-16},
	langid = {english},
	note = {Library Catalog: www.thecreativemomentum.com}
}

@article{mucha_community_2010,
	title = {Community Structure in Time-Dependent, Multiscale, and Multiplex Networks},
	volume = {328},
	issn = {0036-8075, 1095-9203},
	url = {http://arxiv.org/abs/0911.1824},
	doi = {10.1126/science.1184819},
	pages = {876--878},
	number = {5980},
	journaltitle = {Science},
	shortjournal = {Science},
	author = {Mucha, Peter J. and Richardson, Thomas and Macon, Kevin and Porter, Mason A. and Onnela, Jukka-Pekka},
	urldate = {2020-08-30},
	date = {2010-05-14},
	eprinttype = {arxiv},
	eprint = {0911.1824}
}

@inproceedings{dmitriev_dirty_2017,
	location = {New York, {NY}, {USA}},
	title = {A Dirty Dozen: Twelve Common Metric Interpretation Pitfalls in Online Controlled Experiments},
	isbn = {978-1-4503-4887-4},
	url = {http://doi.org/10.1145/3097983.3098024},
	doi = {10.1145/3097983.3098024},
	series = {{KDD} '17},
	shorttitle = {A Dirty Dozen},
	pages = {1427--1436},
	booktitle = {Proceedings of the 23rd {ACM} {SIGKDD} International Conference on Knowledge Discovery and Data Mining},
	publisher = {Association for Computing Machinery},
	author = {Dmitriev, Pavel and Gupta, Somit and Kim, Dong Woo and Vaz, Garnet},
	urldate = {2020-09-01},
	date = {2017-08-13}
}

@inproceedings{kohavi_online_2012,
	location = {New York, {NY}, {USA}},
	title = {Online controlled experiments: introduction, learnings, and humbling statistics},
	isbn = {978-1-4503-1270-7},
	url = {http://doi.org/10.1145/2365952.2365954},
	doi = {10.1145/2365952.2365954},
	series = {{RecSys} '12},
	shorttitle = {Online controlled experiments},
	pages = {1--2},
	booktitle = {Proceedings of the sixth {ACM} conference on Recommender systems},
	publisher = {Association for Computing Machinery},
	author = {Kohavi, Ron},
	urldate = {2020-09-03},
	date = {2012-09-09}
}

@article{parker_optimal_2017,
	title = {Optimal design of experiments on connected units with application to social networks},
	volume = {66},
	rights = {© 2016 Royal Statistical Society},
	issn = {1467-9876},
	url = {http://rss.onlinelibrary.wiley.com/doi/abs/10.1111/rssc.12170},
	doi = {10.1111/rssc.12170},
	pages = {455--480},
	number = {3},
	journaltitle = {Journal of the Royal Statistical Society: Series C (Applied Statistics)},
	author = {Parker, Ben M. and Gilmour, Steven G. and Schormans, John},
	urldate = {2020-09-10},
	date = {2017},
	langid = {english},
	note = {\_eprint: https://onlinelibrary.wiley.com/doi/pdf/10.1111/rssc.12170}
}

@inproceedings{xu_sqr_2018,
	location = {London United Kingdom},
	title = {{SQR}: Balancing Speed, Quality and Risk in Online Experiments},
	isbn = {978-1-4503-5552-0},
	url = {https://dl.acm.org/doi/10.1145/3219819.3219875},
	doi = {10.1145/3219819.3219875},
	shorttitle = {{SQR}},
	eventtitle = {{KDD} '18: The 24th {ACM} {SIGKDD} International Conference on Knowledge Discovery and Data Mining},
	pages = {895--904},
	booktitle = {Proceedings of the 24th {ACM} {SIGKDD} International Conference on Knowledge Discovery \& Data Mining},
	publisher = {{ACM}},
	author = {Xu, Ya and Duan, Weitao and Huang, Shaochen},
	urldate = {2020-10-20},
	date = {2018-07-19},
	langid = {english}
}

@inproceedings{lomas_interface_2016,
	location = {New York, {NY}, {USA}},
	title = {Interface Design Optimization as a Multi-Armed Bandit Problem},
	isbn = {978-1-4503-3362-7},
	url = {http://doi.org/10.1145/2858036.2858425},
	doi = {10.1145/2858036.2858425},
	series = {{CHI} '16},
	pages = {4142--4153},
	booktitle = {Proceedings of the 2016 {CHI} Conference on Human Factors in Computing Systems},
	publisher = {Association for Computing Machinery},
	author = {Lomas, J. Derek and Forlizzi, Jodi and Poonwala, Nikhil and Patel, Nirmal and Shodhan, Sharan and Patel, Kishan and Koedinger, Ken and Brunskill, Emma},
	urldate = {2020-10-26},
	date = {2016-05-07}
}

@inproceedings{li_contextual-bandit_2010,
	location = {New York, {NY}, {USA}},
	title = {A contextual-bandit approach to personalized news article recommendation},
	isbn = {978-1-60558-799-8},
	url = {http://doi.org/10.1145/1772690.1772758},
	doi = {10.1145/1772690.1772758},
	series = {{WWW} '10},
	pages = {661--670},
	booktitle = {Proceedings of the 19th international conference on World wide web},
	publisher = {Association for Computing Machinery},
	author = {Li, Lihong and Chu, Wei and Langford, John and Schapire, Robert E.},
	urldate = {2020-10-27},
	date = {2010-04-26}
}

@article{issa_mattos_multi-armed_2019,
	title = {Multi-armed bandits in the wild: Pitfalls and strategies in online experiments},
	volume = {113},
	issn = {0950-5849},
	url = {http://www.sciencedirect.com/science/article/pii/S0950584919301053},
	doi = {10.1016/j.infsof.2019.05.004},
	shorttitle = {Multi-armed bandits in the wild},
	pages = {68--81},
	journaltitle = {Information and Software Technology},
	shortjournal = {Information and Software Technology},
	author = {Issa Mattos, David and Bosch, Jan and Olsson, Helena Holmström},
	urldate = {2020-11-04},
	date = {2019-09-01},
	langid = {english}
}

@inproceedings{backstrom_network_2011,
	location = {New York, {NY}, {USA}},
	title = {Network bucket testing},
	isbn = {978-1-4503-0632-4},
	url = {http://doi.org/10.1145/1963405.1963492},
	doi = {10.1145/1963405.1963492},
	series = {{WWW} '11},
	pages = {615--624},
	booktitle = {Proceedings of the 20th international conference on World wide web},
	publisher = {Association for Computing Machinery},
	author = {Backstrom, Lars and Kleinberg, Jon},
	urldate = {2020-12-04},
	date = {2011-03-28}
}

@inproceedings{katzir_framework_2012,
	location = {New York, {NY}, {USA}},
	title = {Framework and algorithms for network bucket testing},
	isbn = {978-1-4503-1229-5},
	url = {http://doi.org/10.1145/2187836.2187974},
	doi = {10.1145/2187836.2187974},
	series = {{WWW} '12},
	pages = {1029--1036},
	booktitle = {Proceedings of the 21st international conference on World Wide Web},
	publisher = {Association for Computing Machinery},
	author = {Katzir, Liran and Liberty, Edo and Somekh, Oren},
	urldate = {2020-12-04},
	date = {2012-04-16}
}

@article{newman_modularity_2006,
	title = {Modularity and community structure in networks},
	volume = {103},
	rights = {© 2006 by The National Academy of Sciences of the {USA}},
	issn = {0027-8424, 1091-6490},
	url = {https://www.pnas.org/content/103/23/8577},
	doi = {10.1073/pnas.0601602103},
	pages = {8577--8582},
	number = {23},
	journaltitle = {Proceedings of the National Academy of Sciences},
	shortjournal = {{PNAS}},
	author = {Newman, M. E. J.},
	urldate = {2020-12-07},
	date = {2006-06-06},
	langid = {english},
	pmid = {16723398},
	note = {Publisher: National Academy of Sciences
Section: Physical Sciences}
}

@inproceedings{leskovec_empirical_2010,
	location = {New York, {NY}, {USA}},
	title = {Empirical comparison of algorithms for network community detection},
	isbn = {978-1-60558-799-8},
	url = {http://doi.org/10.1145/1772690.1772755},
	doi = {10.1145/1772690.1772755},
	series = {{WWW} '10},
	pages = {631--640},
	booktitle = {Proceedings of the 19th international conference on World wide web},
	publisher = {Association for Computing Machinery},
	author = {Leskovec, Jure and Lang, Kevin J. and Mahoney, Michael},
	urldate = {2020-12-07},
	date = {2010-04-26}
}

@article{stanley_clustering_2016,
	title = {Clustering Network Layers with the Strata Multilayer Stochastic Block Model},
	volume = {3},
	issn = {2327-4697},
	doi = {10.1109/TNSE.2016.2537545},
	pages = {95--105},
	number = {2},
	journaltitle = {{IEEE} Transactions on Network Science and Engineering},
	author = {Stanley, N. and Shai, S. and Taylor, D. and Mucha, P. J.},
	date = {2016-04},
	note = {Conference Name: {IEEE} Transactions on Network Science and Engineering}
}

@article{zhou_cluster-adaptive_2020,
	title = {Cluster-Adaptive Network A/B Testing: From Randomization to Estimation},
	url = {http://arxiv.org/abs/2008.08648},
	shorttitle = {Cluster-Adaptive Network A/B Testing},
	journaltitle = {{arXiv}:2008.08648 [stat]},
	author = {Zhou, Yifan and Liu, Yang and Li, Ping and Hu, Feifang},
	urldate = {2020-12-08},
	date = {2020-08-19},
	eprinttype = {arxiv},
	eprint = {2008.08648}
}

@article{zhang_optimal_2020,
	title = {Optimal Design for A/B Testing in the Presence of Covariates and Network Connection},
	url = {http://arxiv.org/abs/2008.06476},
	journaltitle = {{arXiv}:2008.06476 [stat]},
	author = {Zhang, Qiong and Kang, Lulu},
	urldate = {2020-12-08},
	date = {2020-08-14},
	eprinttype = {arxiv},
	eprint = {2008.06476}
}

@thesis{koutra_designing_2017,
	title = {Designing experiments on networks},
	rights = {uos\_thesis},
	url = {https://eprints.soton.ac.uk/416580/},
	pagetotal = {222},
	institution = {University of Southampton},
	type = {phdthesis},
	author = {Koutra, Vasiliki},
	editora = {Koutra, Vasiliki and Gilmour, Steven and Parker, Benjamin and Smith, Peter W. F.},
	editoratype = {collaborator},
	urldate = {2020-12-08},
	date = {2017-09},
	langid = {english}
}

@misc{Hern_why_google,
author = {Hern, Alex},
title = {Why Google has 200m reasons to put engineers over designers | Google | The Guardian},
howpublished = {\url{https://www.theguardian.com/technology/2014/feb/05/why-google-engineers-designers}},
month = {2},
year = {2014},
note = {(Accessed on 10/26/2021)}
}

@misc{all_about_ab_testing,
author = {Steve Urban and Rangarajan Sreenivasan and Vineet Kannan},
title = {It’s All A/Bout Testing: The Netflix Experimentation Platform | by Netflix Technology Blog | Netflix TechBlog},
howpublished = {\url{https://netflixtechblog.com/its-all-a-bout-testing-the-netflix-experimentation-platform-4e1ca458c15}},
month = {4},
year = {2016},
note = {(Accessed on 10/26/2021)}
}

@misc{bump_2019, title={Analysis | '60 Minutes' profiles the genius who won Trump's campaign: Facebook}, url={https://www.washingtonpost.com/news/politics/wp/2017/10/09/60-minutes-profiles-the-genius-who-won-trumps-campaign-facebook/}, journal={The Washington Post}, publisher={WP Company}, author={Bump, Philip}, year={2019}, month={3}}

@misc{isaac_2021, title={Facebook Wrestles With the Features It Used to Define Social Networking}, url={https://www.nytimes.com/2021/10/25/technology/facebook-like-share-buttons.html}, journal={The New York Times}, publisher={The New York Times}, author={Isaac, Mike}, year={2021}, month={10}}

@article{austrian2021applying,
  title={Applying A/B Testing to Clinical Decision Support: Rapid Randomized Controlled Trials},
  author={Austrian, Jonathan and Mendoza, Felicia and Szerencsy, Adam and Fenelon, Lucille and Horwitz, Leora I and Jones, Simon and Kuznetsova, Masha and Mann, Devin M},
  journal={Journal of medical Internet research},
  volume={23},
  number={4},
  pages={e16651},
  year={2021},
  publisher={JMIR Publications Inc., Toronto, Canada}
}

@misc{inhouse_exp_platform,
author = {Visser, Denise},
title = {In-house experimentation platforms},
howpublished = {\url{https://www.linkedin.com/pulse/in-house-experimentation-platforms-denise-visser/}},
month = {12},
year = {2020},
note = {(Accessed on 09/15/2022)}
}

@misc{microsof_exp_platform,
author = {Kohlmeier, Sebastian},
title = {Microsoft’s Experimentation Platform: How We Build a World Class Product - Microsoft Research},
howpublished = {\url{https://www.microsoft.com/en-us/research/group/experimentation-platform-exp/articles/microsofts-experimentation-platform-how-we-build-a-world-class-product/}},
month = {1},
year = {2022},
note = {(Accessed on 02/14/2022)}
}

@misc{linkedin_exp_platform,
author = {Ivaniuk, Alexander},
title = {Our evolution towards T-REX: The prehistory of experimentation infrastructure at LinkedIn | LinkedIn Engineering},
howpublished = {\url{https://engineering.linkedin.com/blog/2020/our-evolution-towards-t-rex--the-prehistory-of-experimentation-i}},
month = {9},
year = {2020},
note = {(Accessed on 02/14/2022)}
}

@misc{metric_comp2020,
author={Boucher, Craig and Knoblich, Ulf and Miller, Dan and Patotski, Sasha and Saied, Amin},
title = {Metric Computation for Multiple Backends},
howpublished = {\url{https://www.microsoft.com/en-us/research/group/experimentation-platform-exp/articles/metric-computation-for-multiple-backends/}},
month = {12},
year = {2020},
note = {(Accessed on 09/16/2022)}
}

@book{thomke2020experimentation,
  title={Experimentation works: The surprising power of business experiments},
  author={Thomke, Stefan H},
  year={2020},
  publisher={Harvard Business Press}
}

@book{georgiev2019statmethods,
  title={Statistical methods in online A/B testing},
  author={Georgiev, Georgi Z.},
  year={2019},
  publisher={Self-Published}
}

@inproceedings{karrer2021network,
  title={Network experimentation at scale},
  author={Karrer, Brian and Shi, Liang and Bhole, Monica and Goldman, Matt and Palmer, Tyrone and Gelman, Charlie and Konutgan, Mikael and Sun, Feng},
  booktitle={Proceedings of the 27th ACM SIGKDD Conference on Knowledge Discovery \& Data Mining},
  pages={3106--3116},
  year={2021}
}

@misc{linked_network_abtest_example,
author = {Saint-Jacques, Guillaume},
title = {Detecting interference: An A/B test of A/B tests | LinkedIn Engineering},
howpublished = {\url{https://engineering.linkedin.com/blog/2019/06/detecting-interference--an-a-b-test-of-a-b-tests}},
month = {6},
year = {2019},
note = {(Accessed on 02/22/2022)}
}

@misc{CUPED_at_BBC,
author = {Hopkins, Frank},
title = {Increasing experimental power with variance reduction at the BBC | by Frank Hopkins | BBC Data Science | Medium},
howpublished = {\url{https://medium.com/bbc-data-science/increasing-experiment-sensitivity-through-pre-experiment-variance-reduction-166d7d00d8fd}},
month = {9},
year = {2020},
note = {(Accessed on 02/25/2022)}
}

@misc{evidence_of_CUPED_at_airbnb,
author = {Sharma, Chetan},
title = {Reducing Experiment Durations - Eppo Blog},
howpublished = {\url{https://www.geteppo.com/blog/reducing-experiment-durations}},
month = {6},
year = {2021},
note = {(Accessed on 02/25/2022)}
}

@inproceedings{deng2023zero,
  title={Zero to Hero: Exploiting Null Effects to Achieve Variance Reduction in Experiments with One-sided Triggering},
  author={Deng, Alex and Yuan, Lo-Hua and Kanai, Naoya and Salama-Manteau, Alexandre},
  booktitle={Proceedings of the Sixteenth ACM International Conference on Web Search and Data Mining},
  pages={823--831},
  year={2023}
}

@article{zhang2021regression,
  title={Regression Adjustment with Synthetic Controls in Online Experiments},
  author={Zhang, Congshan and Coey, Dominic and Goldman, Matt and Karrer, Brian},
  year={2021}
}

@INPROCEEDINGS{fabijan_exp_platform_survey,
  author={Fabijan, Aleksander and Dmitriev, Pavel and Holmstrom Olsson, Helena and Bosch, Jan},
  booktitle={2018 44th Euromicro Conference on Software Engineering and Advanced Applications (SEAA)}, 
  title={Online Controlled Experimentation at Scale: An Empirical Survey on the Current State of A/B Testing}, 
  year={2018},
  volume={},
  number={},
  pages={68-72},
  doi={10.1109/SEAA.2018.00021}
}

@article{petersen2016fused,
  title={Fused lasso additive model},
  author={Petersen, Ashley and Witten, Daniela and Simon, Noah},
  journal={Journal of Computational and Graphical Statistics},
  volume={25},
  number={4},
  pages={1005--1025},
  year={2016},
  publisher={Taylor \& Francis}
}

@article{peysakhovich2016combining,
  title={Combining observational and experimental data to find heterogeneous treatment effects},
  author={Peysakhovich, Alexander and Lada, Akos},
  journal={arXiv preprint arXiv:1611.02385},
  year={2016}
}

@misc{netflix_proxy_metrics,
author = {Gibson Biddle},
title = {Proxy Metrics: How to define a metric to prove or disprove your hypotheses and measure progress},
howpublished = {\url{https://gibsonbiddle.medium.com/4-proxy-metrics-a82dd30ca810}},
month = {7},
year = {2019},
note = {(Accessed on 03/04/2022)}
}

@techreport{athey2019surrogate,
  title={The surrogate index: Combining short-term proxies to estimate long-term treatment effects more rapidly and precisely},
  author={Athey, Susan and Chetty, Raj and Imbens, Guido W and Kang, Hyunseung},
  year={2019},
  institution={National Bureau of Economic Research}
}

@article{pramanik2021modified,
  title={A modified sequential probability ratio test},
  author={Pramanik, Sandipan and Johnson, Valen E and Bhattacharya, Anirban},
  journal={Journal of Mathematical Psychology},
  volume={101},
  pages={102505},
  year={2021},
  publisher={Elsevier}
}

@misc{optimizely_alwaysvalid,
author = {Leonid Pekelis},
title = {Statistics for the Internet Age: The Story Behind Optimizely’s New Stats Engine},
howpublished = {\url{https://www.optimizely.com/insights/blog/statistics-for-the-internet-age-the-story-behind-optimizelys-new-stats-engine/}},
month = {1},
year = {2015},
note = {(Accessed on 03/08/2022)}
}

@misc{Forbes_data_analytics,
author = {Schroeder, Bernhard},
title = {The Data Analytics Profession And Employment Is Exploding: Three Trends That Matter},
howpublished = {\url{https://www.forbes.com/sites/bernhardschroeder/2021/06/11/the-data-analytics-profession-and-employment-is-exploding-three-trends-that-matter/?sh=12c5c3c3f81e}},
month = {6},
year = {2021},
note = {(Accessed on 03/10/2022)}
}

@article{stevens2022comparative,
  title={Comparative Probability Metrics: Using Posterior Probabilities to Account for Practical Equivalence in A/B tests},
  author={Stevens, Nathaniel T and Hagar, Luke},
  journal={The American Statistician},
  volume={76},
  number={3},
  pages={224--237},
  year={2022},
  publisher={Taylor \& Francis}
}

@incollection{kamalbasha2021bayesian,
  title={Bayesian A/B testing for business decisions},
  author={Kamalbasha, Shafi and Eugster, Manuel JA},
  booktitle={Data science--analytics and applications},
  pages={50--57},
  year={2021},
  publisher={Springer}
}

@article{hoffmann2021bayesian,
  title={Bayesian inference for the A/B test: Example applications with r and jasp},
  author={Hoffmann, Tabea and Wagenmakers, Eric-Jan},
  journal={PsyArXiv. June},
  volume={10},
  year={2021}
}

@article{liu2019large,
  title={Large-scale online experimentation with quantile metrics},
  author={Liu, Min and Sun, Xiaohui and Varshney, Maneesh and Xu, Ya},
  journal={arXiv preprint arXiv:1903.08762},
  year={2019}
}

@article{howard2019sequential,
  title={Sequential estimation of quantiles with applications to A/B-testing and best-arm identification},
  author={Howard, Steven R and Ramdas, Aaditya},
  journal={arXiv preprint arXiv:1906.09712},
  year={2019}
}

@misc{uber_quantiles_2018,
author = {Lux, Matthias},
title = {Analyzing Experiment Outcomes: Beyond Average Treatment Effects - Uber Engineering Blog},
howpublished = {\url{https://eng.uber.com/analyzing-experiment-outcomes/}},
month = {11},
year = {2018},
note = {(Accessed on 03/21/2022)}
}

@misc{overlapping_2021,
author = {Chan, Timothy},
title = {Embrace Overlapping A/B Tests and Avoid the Dangers of Isolating Experiments},
howpublished = {\url{https://blog.statsig.com/embracing-overlapping-a-b-tests-and-the-danger-of-isolating-experiments-cb0a69e09d3}},
month = {10},
year = {2021},
note = {(Accessed on 03/21/2022)}
}

@article{bojinov2021panel,
  title={Panel experiments and dynamic causal effects: A finite population perspective},
  author={Bojinov, Iavor and Rambachan, Ashesh and Shephard, Neil},
  journal={Quantitative Economics},
  volume={12},
  number={4},
  pages={1171--1196},
  year={2021},
  publisher={Wiley Online Library}
}

@article{dimakopoulou2021online,
  title={Online Multi-Armed Bandits with Adaptive Inference},
  author={Dimakopoulou, Maria and Ren, Zhimei and Zhou, Zhengyuan},
  journal={Advances in Neural Information Processing Systems},
  volume={34},
  year={2021}
}

@misc{stitch_fix_mabs,
author = {Amadio, Brian},
title = {Multi-Armed Bandits and the Stitch Fix Experimentation Platform},
howpublished = {\url{https://multithreaded.stitchfix.com/blog/2020/08/05/bandits/}},
month = {8},
year = {2020},
note = {(Accessed on 03/21/2022)}
}

@misc{HowGoogl12:online,
author = {{Google}},
title = {How Google's Algorithm is Focused on Its Users - Google Search},
howpublished = {\url{https://www.google.com/search/howsearchworks/mission/users/}},
month = {},
year = {2022},
note = {(Accessed on 03/29/2022)}
}

@article{bajari2021multiple,
  title={Multiple Randomization Designs},
  author={Bajari, Patrick and Burdick, Brian and Imbens, Guido W and Masoero, Lorenzo and McQueen, James and Richardson, Thomas and Rosen, Ido M},
  journal={arXiv preprint arXiv:2112.13495},
  year={2021}
}

@book{box_hunter_hunter_2005, place={Hoboken, New Jersey}, edition={2}, title={Statistics for Experimenters: Design, Innovation, and Discovery}, publisher={Wiley-Interscience}, author={Box, George E.P. and Hunter, J Stuart and Hunter, William G}, year={2005}}

@inproceedings{kohaviIntuitionBusters,
author = {Kohavi, Ron and Deng, Alex and Vermeer, Lukas},
year = {2022},
month = {08},
pages = {},
title = {A/B Testing Intuition Busters: Common Misunderstandings in Online Controlled Experiments},
doi = {10.1145/3534678.3539160}
}

@misc{lindon_anytime_2020,
  doi = {10.48550/ARXIV.2011.03567},
  
  url = {https://arxiv.org/abs/2011.03567},
  
  author = {Lindon, Michael and Malek, Alan},
  
  title = {Anytime-Valid Inference for Multinomial Count Data},
  
  publisher = {arXiv},
  
  year = {2020},
  
  copyright = {arXiv.org perpetual, non-exclusive license}
}

@inproceedings{liu2021trustworthy,
  title={Trustworthy and Powerful Online Marketplace Experimentation with Budget-split Design},
  author={Liu, Min and Mao, Jialiang and Kang, Kang},
  booktitle={Proceedings of the 27th ACM SIGKDD Conference on Knowledge Discovery \& Data Mining},
  pages={3319--3329},
  year={2021}
}

@article{holtz2020reducing,
  title={Reducing interference bias in online marketplace pricing experiments},
  author={Holtz, David and Lobel, Ruben and Liskovich, Inessa and Aral, Sinan},
  journal={arXiv preprint arXiv:2004.12489},
  year={2020}
}

@misc{Bending_time,
author = {Chetan Sharma},
title = {Bending time in experimentation - Eppo Blog},
howpublished = {\url{https://www.geteppo.com/blog/bending-time-in-experimentation}},
month = {6},
year = {2022},
note = {(Accessed on 08/18/2022)}
}

@misc{cuped_statsig,
author = {Craig Sexauer},
title = {CUPED on Statsig},
howpublished = {\url{https://blog.statsig.com/cuped-on-statsig-d57f23122d0e}},
month = {},
year = {2022},
note = {(Accessed on 08/18/2022)}
}

@misc{cuped_towardsdatascience,
author = {Matteo Courthoud},
title = {Understanding CUPED},
howpublished = {\url{https://towardsdatascience.com/understanding-cuped-a822523641af}},
month = {6},
year = {2022},
note = {(Accessed on 08/18/2022)}
}

@misc{netflix_dml,
author = {Yinghong Lan and Vinod Bakthavachalam and Lavanya Sharan and Marie Douriez and Bahar Azarnoush and Mason Kroll},
title = {A Survey of Causal Inference Applications at Netflix | by Netflix Technology Blog},
howpublished = {\url{https://netflixtechblog.com/a-survey-of-causal-inference-applications-at-netflix-b62d25175e6f}},
month = {5},
year = {2022},
note = {(Accessed on 08/18/2022)}
}

@misc{athey2020combine,
  doi = {10.48550/arxiv.2006.09676},
  url = {https://arxiv.org/abs/2006.09676},
  author = {Athey, Susan and Chetty, Raj and Imbens, Guido},
  title = {Combining Experimental and Observational Data to Estimate Treatment Effects on Long Term Outcomes},
  publisher = {arXiv},
  year = {2020},
  copyright = {arXiv.org perpetual, non-exclusive license}
}

@misc{imbens2022longterm,
  doi = {10.48550/ARXIV.2202.07234},
  url = {https://arxiv.org/abs/2202.07234},
  author = {Imbens, Guido and Kallus, Nathan and Mao, Xiaojie and Wang, Yuhao},
  title = {Long-term Causal Inference Under Persistent Confounding via Data Combination},
  publisher = {arXiv},
  year = {2022},
  copyright = {arXiv.org perpetual, non-exclusive license}
}

@article{berman2021false,
  title={False discovery in A/B testing},
  author={Berman, Ron and Van den Bulte, Christophe},
  journal={Management Science},
  year={2021},
  publisher={INFORMS}
}

@inproceedings{blake2014marketplace,
  title={Why marketplace experimentation is harder than it seems: The role of test-control interference},
  author={Blake, Thomas and Coey, Dominic},
  booktitle={Proceedings of the fifteenth ACM conference on Economics and computation},
  pages={567--582},
  year={2014}
}

@misc{basic2019uncontrolled,
  title={UNCONTROLLED The Surprising Payoff of Trial-and-Error for Business, Politics, and Society},
  author={Manzi, Jim},
  year={2012},
  publisher={Basic Books}
}

@article{koning2022experimentation,
  title={Experimentation and Start-up Performance: Evidence from A/B Testing},
  author={Koning, Rembrand and Hasan, Sharique and Chatterji, Aaron},
  journal={Management Science},
  year={2022},
  publisher={INFORMS}
}

@article{athey2018exact,
  title={Exact p-values for network interference},
  author={Athey, Susan and Eckles, Dean and Imbens, Guido W},
  journal={Journal of the American Statistical Association},
  volume={113},
  number={521},
  pages={230--240},
  year={2018},
  publisher={Taylor \& Francis}
}

@book{luca2021power,
  title={The power of experiments: Decision making in a data-driven world},
  author={Luca, Michael and Bazerman, Max H},
  year={2021},
  publisher={Mit Press}
}

@misc{kohavi2023slides,
    title = {Build vs Buy},
    author = {Kohavi, Ronny},
    year = {2023},
    howpublished = {\url{https://bit.ly/ABTestsBuildVsBuy8}}
}

@inproceedings{lindon2022rapid,
author = {Lindon, Michael and Sanden, Chris and Shirikian, Vach\'{e}},
title = {Rapid Regression Detection in Software Deployments through Sequential Testing},
year = {2022},
isbn = {9781450393850},
publisher = {Association for Computing Machinery},
address = {New York, NY, USA},
url = {https://doi.org/10.1145/3534678.3539099},
doi = {10.1145/3534678.3539099},
booktitle = {Proceedings of the 28th ACM SIGKDD Conference on Knowledge Discovery and Data Mining},
pages = {3336–3346},
numpages = {11},
location = {Washington DC, USA},
series = {KDD '22}
}

@article{benbunan2017ethics,
  title={The ethics of online research with unsuspecting users: From A/B testing to C/D experimentation},
  author={Benbunan-Fich, Raquel},
  journal={Research Ethics},
  volume={13},
  number={3-4},
  pages={200--218},
  year={2017},
  publisher={SAGE Publications Sage UK: London, England}
}

@misc{convert2021ethics,
    title = {The Guide to Ethical A/B Testing: The Missing Component of Your Optimization Program},
    author = {Kontotasiou, Dionysia},
    year = {2021},
    howpublished = {\url{convert.com/blog/a-b-testing/ethical-ab-testing-guide/}}
}

@article{kramer2014experimental,
  title={Experimental evidence of massive-scale emotional contagion through social networks},
  author={Kramer, Adam DI and Guillory, Jamie E and Hancock, Jeffrey T},
  journal={Proceedings of the National Academy of Sciences},
  volume={111},
  number={24},
  pages={8788--8790},
  year={2014},
  publisher={National Acad Sciences}
}

@misc{rudder2014okcupid,
    title = {We experiment on human beings!},
    author = {Rudder, Christian},
    year = {2014},
    howpublished = {\url{https://www.gwern.net/docs/psychology/okcupid/weexperimentonhumanbeings.html}}
}

@article{rajkumar2022causal,
  title={A causal test of the strength of weak ties},
  author={Rajkumar, Karthik and Saint-Jacques, Guillaume and Bojinov, Iavor and Brynjolfsson, Erik and Aral, Sinan},
  journal={Science},
  volume={377},
  number={6612},
  pages={1304--1310},
  year={2022},
  publisher={American Association for the Advancement of Science}
}

@book{wald1947sequential,
  title={Sequential analysis},
  author={Wald, Abraham},
  year={1947},
  publisher={Courier Corporation}
}

@misc{georgi2022gst,
    title = {Fully Sequential vs Group Sequential Test},
    author = {Georgiev, Georgi},
    year = {2022},
    howpublished = {\url{https://blog.analytics-toolkit.com/2022/fully-sequential-vs-group-sequential-tests}}
}

@article{bojinov2022online,
  title={Online experimentation: Benefits, operational and methodological challenges, and scaling guide},
  author={Bojinov, Iavor and Gupta, Somit},
  journal={Harvard Data Science Review},
  volume={4},
  number={3},
  year={2022}
}

@misc{netflix2017interleave,
    title = {Innovating Faster on Personalization Algorithms at Netflix Using Interleaving},
    author = {Parks, Joshua and Aurisset, Juliette and Ramm, Michael},
    year = {2017},
    howpublished = {\url{https://netflixtechblog.com/interleaving-in-online-experiments-at-netflix-a04ee392ec55}}
}

@article{sadeghi2022novelty,
  title={Novelty and primacy: a long-term estimator for online experiments},
  author={Sadeghi, Soheil and Gupta, Somit and Gramatovici, Stefan and Lu, Jiannan and Ai, Hao and Zhang, Ruhan},
  journal={Technometrics},
  volume={64},
  number={4},
  pages={524--534},
  year={2022},
  publisher={Taylor \& Francis}
}

@inproceedings{microsoft2013interleaving,
author = {Radlinski, Filip and Craswell, Nick},
title = {Optimized Interleaving for Online Retrieval Evaluation},
year = {2013},
isbn = {9781450318693},
publisher = {Association for Computing Machinery},
address = {New York, NY, USA},
url = {https://doi.org/10.1145/2433396.2433429},
doi = {10.1145/2433396.2433429},
booktitle = {Proceedings of the Sixth ACM International Conference on Web Search and Data Mining},
pages = {245–254},
numpages = {10},
location = {Rome, Italy},
series = {WSDM '13}
}

@misc{airbnb2022interleave,
    title = {Beyond A/B Test: Speeding up Airbnb Search Ranking Experimentation through Interleaving},
    author = {Zhang, Qing and Du, Michelle and Andersen, Reid and Hel, Liwei},
    year = {2022},
    howpublished = {\url{https://medium.com/airbnb-engineering/beyond-a-b-test-speeding-up-airbnb-search-ranking-experimentation-through-interleaving-7087afa09c8e}}
}

@article{tsiatis2006semiparametric,
  title={Semiparametric theory and missing data},
  author={Tsiatis, Anastasios A},
  year={2006},
  publisher={Springer}
}

@misc{duolingo2022testeverything,
    title = {Shareholder Letter Q2 2022},
    author = {Von Ahn, Luis},
    year = {2022},
    howpublished = {\url{https://investors.duolingo.com/static-files/ae55dd31-2ce4-41ac-bb26-948bafe8409c}}
}

@article{neyman1923application,
  title={On the application of probability theory to agricultural experiments. Essay on principles},
  author={Neyman, Jerzy},
  journal={Ann. Agricultural Sciences},
  pages={1--51},
  year={1923}
}

@article{rubin1974estimating,
  title={Estimating causal effects of treatments in randomized and nonrandomized studies.},
  author={Rubin, Donald B},
  journal={Journal of educational Psychology},
  volume={66},
  number={5},
  pages={688},
  year={1974},
  publisher={American Psychological Association}
}

@article{mcfowland2021prescriptive,
  title={A prescriptive analytics framework for optimal policy deployment using heterogeneous treatment effects.},
  author={McFowland III, Edward and Gangarapu, Sandeep and Bapna, Ravi and Sun, Tianshu},
  journal={MIS Quarterly},
  volume={45},
  number={4},
  year={2021}
}

@article{sepehri2020interpretable,
  title={Interpretable Assessment of Fairness During Model Evaluation},
  author={Sepehri, Amir and DiCiccio, Cyrus},
  journal={arXiv preprint arXiv:2010.13782},
  year={2020}
}

@article{waudby2021time,
  title={Time-uniform central limit theory, asymptotic confidence sequences, and anytime-valid causal inference},
  author={Waudby-Smith, Ian and Arbour, David and Sinha, Ritwik and Kennedy, Edward H and Ramdas, Aaditya},
  journal={arXiv preprint arXiv:2103.06476},
  year={2021}
}

@article{ham2022design,
  title={Design-Based Confidence Sequences for Anytime-valid Causal Inference},
  author={Ham, Dae Woong and Bojinov, Iavor and Lindon, Michael and Tingley, Martin},
  journal={arXiv preprint arXiv:2210.08639},
  year={2022}
}

@article{matias2021upworthy,
  title={The Upworthy Research Archive, a time series of 32,487 experiments in US media},
  author={Matias, J Nathan and Munger, Kevin and Le Quere, Marianne Aubin and Ebersole, Charles},
  journal={Scientific Data},
  volume={8},
  number={1},
  pages={195},
  year={2021},
  publisher={Nature Publishing Group UK London}
}

@article{liu2021datasets,
  title={Datasets for online controlled experiments},
  author={Liu, CH and Cardoso, {\^A}ngelo and Couturier, Paul and McCoy, Emma J},
  journal={arXiv preprint arXiv:2111.10198},
  year={2021}
}

@article{basse2023minimax,
  title={Minimax designs for causal effects in temporal experiments with treatment habituation},
  author={Basse, Guillaume W and Ding, Yi and Toulis, Panos},
  journal={Biometrika},
  volume={110},
  number={1},
  pages={155--168},
  year={2023},
  publisher={Oxford University Press}
}

@article{bojinov2022design,
  title={Design and analysis of switchback experiments},
  author={Bojinov, Iavor and Simchi-Levi, David and Zhao, Jinglong},
  journal={Management Science},
  year={2022},
  publisher={INFORMS}
}

@inproceedings{miroglio2018effect,
  title={The effect of ad blocking on user engagement with the web},
  author={Miroglio, Ben and Zeber, David and Kaye, Jofish and Weiss, Rebecca},
  booktitle={Proceedings of the 2018 world wide web conference},
  pages={813--821},
  year={2018}
}

@misc{netflix2018quasi,
    title = {Quasi Experimentation at Netflix},
    author = {McFarland, Colin and Pow, Michael and Glick, Julia},
    year = {2018},
    howpublished = {\url{https://netflixtechblog.com/quasi-experimentation-at-netflix-566b57d2e362}}
}

@misc{harinen2019using,
    title = {Using causal inference to improve the uber user experience},
    author = {Harinen, Totte and Li, Bonnie},
    year = {2019},
    howpublished = {\url{https://www.uber.com/en-CA/blog/causal-inference-at-uber/}}
}

@misc{spotify2023sequential,
    title = {Choosing Sequential Testing Framework — Comparisons and Discussions},
    author = {Schultzberg, Mårten and Ankargren, Sebastian},
    year = {2023},
    howpublished = {\url{https://engineering.atspotify.com/2023/03/choosing-sequential-testing-framework-comparisons-and-discussions/}}
}

@Inbook{Kohavi2023,
author="Kohavi, Ron
and Longbotham, Roger",
editor="Phung, Dinh
and Webb, Geoffrey I.
and Sammut, Claude",
title="Online Controlled Experiments and A/B Tests",
bookTitle="Encyclopedia of Machine Learning and Data Science",
year="2023",
publisher="Springer US",
address="New York, NY",
pages="1--13",
isbn="978-1-4899-7502-7",
doi="10.1007/978-1-4899-7502-7_891-2",
url="https://doi.org/10.1007/978-1-4899-7502-7_891-2"
}

@article{rahman2023experimental,
  title={The Experimental Hand: How Platform-based Experimentation Reconfigures Worker Autonomy},
  author={Rahman, Hatim A and Weiss, Tim and Karunakaran, Arvind},
  journal={Academy of Management Journal},
  number={ja},
  year={2023}
}

@article{ruberg1995doseI,
  title={Dose response studies I. Some design considerations},
  author={Ruberg, Stephen J},
  journal={Journal of Biopharmaceutical Statistics},
  volume={5},
  number={1},
  pages={1--14},
  year={1995},
  publisher={Taylor \& Francis}
}

@article{ruberg1995doseII,
  title={Dose response studies II. Analysis and interpretation},
  author={Ruberg, Stephen J},
  journal={Journal of biopharmaceutical statistics},
  volume={5},
  number={1},
  pages={15--42},
  year={1995},
  publisher={Taylor \& Francis}
}

@book{imbens2015causal,
  title={Causal inference in statistics, social, and biomedical sciences},
  author={Imbens, Guido W and Rubin, Donald B},
  year={2015},
  publisher={Cambridge University Press}
}

@article{abadie2020sampling,
  title={Sampling-Based versus Design-Based Uncertainty in Regression Analysis},
  author={Abadie, Alberto and Athey, Susan and Imbens, Guido W and Wooldridge, Jeffrey M},
  journal={Econometrica},
  volume={88},
  number={1},
  pages={265--296},
  year={2020},
  publisher={Wiley Online Library}
}

@misc{lyft2016interference,
    title = {Experimentation in a Ridesharing Marketplace},
    author = {Chamandy, Nicholas},
    year = {2016},
    howpublished = {\url{https://eng.lyft.com/experimentation-in-a-ridesharing-marketplace-b39db027a66e}}
}

@article{tangcontrol,
  title={Control Using Predictions as Covariates in Switchback Experiments},
  author={Tang, Yixin and Huang, Caixia and Kastelman, David and Bauman, Jared},
  year = {2020}
}

@misc{singer2022linkedin,
    title = {LinkedIn Ran Social Experiments on 20 Million Users Over Five Years},
    author = {Singer, Natasha},
    year = {2022},
    howpublished = {\url{https://www.nytimes.com/2022/09/24/business/linkedin-social-experiments.html}}
}

@article{hu2022switchback,
  title={Switchback Experiments under Geometric Mixing},
  author={Hu, Yuchen and Wager, Stefan},
  journal={arXiv preprint arXiv:2209.00197},
  year={2022}
}

@article{ni2023design,
  title={Design of Panel Experiments with Spatial and Temporal Interference},
  author={Ni, Tu and Bojinov, Iavor and Zhao, Jinglong},
  journal={Available at SSRN 4466598},
  year={2023}
}

@article{johari2022experimental,
  title={Experimental design in two-sided platforms: An analysis of bias},
  author={Johari, Ramesh and Li, Hannah and Liskovich, Inessa and Weintraub, Gabriel Y},
  journal={Management Science},
  volume={68},
  number={10},
  pages={7069--7089},
  year={2022},
  publisher={INFORMS}
}

@inproceedings{li2022interference,
  title={Interference, bias, and variance in two-sided marketplace experimentation: Guidance for platforms},
  author={Li, Hannah and Zhao, Geng and Johari, Ramesh and Weintraub, Gabriel Y},
  booktitle={Proceedings of the ACM Web Conference 2022},
  pages={182--192},
  year={2022}
}

@article{bui2023general,
  title={General Additive Network Effect Models},
  author={Bui, Trang and Steiner, Stefan H and Stevens, Nathaniel T},
  journal={The New England Journal of Statistics in Data Science},
  pages={1--19},
  year={2023},
  publisher={New England Statistical Society}
}

@article{parker2017optimal,
  title={Optimal design of experiments on connected units with application to social networks},
  author={Parker, Ben M and Gilmour, Steven G and Schormans, John},
  journal={Journal of the Royal Statistical Society. Series C (Applied Statistics)},
  pages={455--480},
  year={2017},
  publisher={JSTOR}
}

@article{koutra2021optimal,
  title={Optimal block designs for experiments on networks},
  author={Koutra, Vasiliki and Gilmour, Steven G and Parker, Ben M},
  journal={Journal of the Royal Statistical Society Series C: Applied Statistics},
  volume={70},
  number={3},
  pages={596--618},
  year={2021},
  publisher={Oxford University Press}
}

@article{basse2018model,
  title={Model-assisted design of experiments in the presence of network-correlated outcomes},
  author={Basse, Guillaume W and Airoldi, Edoardo M},
  journal={Biometrika},
  volume={105},
  number={4},
  pages={849--858},
  year={2018},
  publisher={Oxford University Press}
}

@article{pokhilko2019d,
  title={D-optimal design for network a/b testing},
  author={Pokhilko, Victoria and Zhang, Qiong and Kang, Lulu and others},
  journal={Journal of Statistical Theory and Practice},
  volume={13},
  number={4},
  pages={1--23},
  year={2019},
  publisher={Springer}
}

@article{zhang2022locally,
  title={Locally Optimal Design for A/B Tests in the Presence of Covariates and Network Dependence},
  author={Zhang, Qiong and Kang, Lulu},
  journal={Technometrics},
  volume={64},
  number={3},
  pages={358--369},
  year={2022},
  publisher={Taylor \& Francis}
}

@misc{booking2023sequential,
    title = {Sequential Testing at Booking.com},
    author = {Skotara, Nils},
    year = {2023},
    howpublished = {\url{https://booking.ai/sequential-testing-at-booking-com-650954a569c7}}
}

@article{robertson2023point,
  title={Point estimation for adaptive trial designs I: A methodological review},
  author={Robertson, David S and Choodari-Oskooei, Babak and Dimairo, Munya and Flight, Laura and Pallmann, Philip and Jaki, Thomas},
  journal={Statistics in medicine},
  volume={42},
  number={2},
  pages={122--145},
  year={2023},
  publisher={Wiley Online Library}
}

@article{pocock1977group,
  title={Group sequential methods in the design and analysis of clinical trials},
  author={Pocock, Stuart J},
  journal={Biometrika},
  volume={64},
  number={2},
  pages={191--199},
  year={1977},
  publisher={Oxford University Press}
}

@article{o1979multiple,
  title={A multiple testing procedure for clinical trials},
  author={O'Brien, Peter C and Fleming, Thomas R},
  journal={Biometrics},
  pages={549--556},
  year={1979},
  publisher={JSTOR}
}

@article{granovetter1973strength,
  title={The strength of weak ties},
  author={Granovetter, Mark S},
  journal={American journal of sociology},
  volume={78},
  number={6},
  pages={1360--1380},
  year={1973},
  publisher={University of Chicago Press}
}

@misc{belanger2022ethics,
    title = {xperts debate the ethics of LinkedIn’s algorithm experiments on 20M users},
    author = {Belanger, Ashley},
    year = {2022},
    howpublished = {\url{https://arstechnica.com/tech-policy/2022/09/experts-debate-the-ethics-of-linkedins-algorithm-experiments-on-20m-users/}}
}

@article{lan1983,
 ISSN = {00063444},
 URL = {http://www.jstor.org/stable/2336502},
 author = {K. K. Gordon Lan and David L. DeMets},
 journal = {Biometrika},
 number = {3},
 pages = {659--663},
 publisher = {[Oxford University Press, Biometrika Trust]},
 title = {Discrete Sequential Boundaries for Clinical Trials},
 urldate = {2023-07-01},
 volume = {70},
 year = {1983}
}

@article{quin2023b,
  title={A/B Testing: A Systematic Literature Review},
  author={Quin, Federico and Weyns, Danny and Galster, Matthias and Silva, Camila Costa},
  journal={arXiv preprint arXiv:2308.04929},
  year={2023}
}

@inproceedings{xia2019safe,
  title={Safe velocity: a practical guide to software deployment at scale using controlled rollout},
  author={Xia, Tong and Bhardwaj, Sumit and Dmitriev, Pavel and Fabijan, Aleksander},
  booktitle={2019 IEEE/ACM 41st International Conference on Software Engineering: Software Engineering in Practice (ICSE-SEIP)},
  pages={11--20},
  year={2019},
  organization={IEEE}
}

\end{document}